\documentclass[aps,twocolumn,showpacs,preprintnumbers,prx,amsmath,amssymb,amsfonts,superscriptaddress,floatfix]{revtex4}

\usepackage[latin1]{inputenc}
\usepackage{graphics}
\usepackage{color}

\usepackage{amssymb}
\usepackage{amsmath}
\usepackage{overpic}
\usepackage{epstopdf}
\usepackage{bm}
\usepackage{graphicx}
\usepackage{subfigure}

\begin{document}

\title{High-pressure Phase Stability and Superconductivity of
  Pnictogen Hydrides and Chemical Trends for Compressed Hydrides}

\author{Yuhao Fu}\thanks{These two authors contributed equally to this work.}
\affiliation{College of Materials Science and Engineering and Key Laboratory of Automobile Materials of MOE, Jilin University, Changchun 130012, China}
\author{Xiangpo Du}\thanks{These two authors contributed equally to this work.}
\affiliation{State Key Laboratory of Superhard Materials, Jilin University, Changchun 130012, China}
\author{Lijun Zhang}
\email{lijun_zhang@jlu.edu.cn}
\affiliation{College of Materials Science and Engineering and Key Laboratory of Automobile Materials of MOE, Jilin University, Changchun 130012, China}
\affiliation{State Key Laboratory of Superhard Materials, Jilin University, Changchun 130012, China}
\author{Feng Peng}
\affiliation{State Key Laboratory of Superhard Materials, Jilin University, Changchun 130012, China}
\author{Miao Zhang}
\affiliation{State Key Laboratory of Superhard Materials, Jilin University, Changchun 130012, China}
\author{Chris J. Pickard}
\affiliation{Department of Materials Science $\&$ Metallurgy, University of Cambridge, 27 Charles Babbage Road, Cambridge CB3 0FS, United Kingdom}
\author{Richard J. Needs}
\affiliation{Theory of Condensed Matter Group, Cavendish Laboratory, J J Thomson Avenue, Cambridge CB3 0HE, United Kingdom}
\author{David J. Singh}
\email{singhdj@missouri.edu}
\affiliation{College of Materials Science and Engineering and Key Laboratory of Automobile Materials of MOE, Jilin University, Changchun 130012, China}
\affiliation{Department of Physics and Astronomy, University of Missouri, Columbia, MO 65211-7010 USA}
\author{Weitao Zheng}
\affiliation{College of Materials Science and Engineering and Key Laboratory of Automobile Materials of MOE, Jilin University, Changchun 130012, China}
\author{Yanming Ma}
\email{mym@jlu.edu.cn}
\affiliation{State Key Laboratory of Superhard Materials, Jilin University, Changchun 130012, China}

\date{\today}

\begin{abstract}
  The recent breakthrough discovery of unprecedentedly high
  temperature superconductivity of 203 K in compressed sulfur hydrides
  has stimulated significant interest in finding new
  hydrogen-containing superconductors and elucidating the physical and
  chemical principles that govern these materials and their
  superconductivity.  Here we report the prediction of high
  temperature superconductivity in the family of pnictogen hydrides
  using first principles calculations in combination with global
  optimization structure searching methods.  The hitherto unknown
  high-pressure phase diagrams of binary hydrides formed by the
  pnictogens of phosphorus, arsenic and antimony are explored, stable
  structures are identified and their electronic, vibrational and
  superconducting properties are investigated.  We predict that
  SbH$_{4}$ and AsH$_{8}$ are high-temperature superconductors at
  megabar pressures, with critical temperatures in excess of 100 K.
  The highly symmetrical hexagonal SbH$_{4}$ phase is predicted to be
  stabilized above about 150 GPa, which is readily achievable in
  diamond anvil cell experiments.  We find that all phosphorus
  hydrides are metastable with respect to decomposition into the
  elements within the pressure range studied.  Trends based on our
  results and data in the literature reveal a connection between the
  high-pressure behaviors and ambient-pressure chemical quantities
  which provides insight into understanding which elements may form
  hydrogen-rich high-temperature superconducting phases at high
  pressures.
\end{abstract}

\maketitle

\section{\textbf{I. INTRODUCTION}}

Superconductivity, especially at high temperatures, continues to offer
surprises both from experimental discoveries and theoretical insights
often stimulated by these experiments.  These findings include the
Fe-based superconductors \cite{jacs.130.3296} which have provided a
large body of theoretical work on spin-fluctuation pairing, nematicity
and multiorbital physics \cite{PhysRevB.77.224509,
  PhysRevLett.101.057003, RevModPhys.84.1383}, and recent reports of
superconductivity in sulfur hydrides at pressures $\sim$200 GPa with a
$T_{c}$ of 203 K \cite{Nature.525.73, arXiv.1509.03156E}.
Superconducting sulfur hydrides show a strong isotope effect when
deuterium is substituted for hydrogen, as expected for
electron-phonon-based superconductivity \cite{Nature.525.73}.  This
breakthrough observation has led to debate on the superconducting
mechanism, including discussion of the limits of standard
Migdal-Eliashberg electron-phonon coupling (EPC) theory
\cite{PhysRevB.91.220507, arXiv.1507.01093B, PhysicaC.511.45}, the
role of strong covalent bonding \cite{PhysRevB.91.060511,
  PhysRevB.92.060508}, the effects of anharmonicity on EPC
\cite{PhysRevLett.114.157004}, $etc$.  These proposals, together with
the pioneering work of Ashcroft and coworkers on predicted
high-temperature superconductivity in solid hydrogen
\cite{PhysRevLett.21.1748} and in hydrogen-rich hydrides
\cite{PhysRevLett.92.187002, PhysRevLett.96.017006} may guide research
towards the discovery of new high-temperature superconductors based on
materials containing light elements.

High-temperature superconductors can be divided into two classes:
unconventional superconductors characterized by non-phonon primary
pairing interactions, normally leading to non-standard pairing states,
and conventional superconductors, with dominant electron-phonon
pairing, and s-wave (or mainly s-wave for non-centrosymmetric cases)
pairing states.  In the latter case, methods exist for reliably
evaluating superconducting properties and analyzing the
superconductivity, \textit{e.g.}, through calculating the Eliashberg
EPC spectral function $\alpha^{2}$F$(\omega)$.  A general consensus
has been reached that the recently discovered superconducting sulfur
hydrides belong to this category \cite{JCP.140.174712, Nature.525.73,
  ScientificReports.4, PhysRevB.91.184511, arXiv.1501.06336F,
  PhysRevB.91.060511}.  The experimental discovery was motivated by a
theoretical prediction of high-$T_{c}$ superconductivity in compressed
solid H$_{2}$S by some of us \cite{JCP.140.174712}.  Two
superconducting states are observed in H$_2$S: samples prepared at low
temperature have a maximum $T_{c}$ of $\sim$ 150 K at 200 GPa, while
the superconductivity at 203 K arises from samples prepared via
annealing at temperatures above 220 K.  The latter phase originates
from the decomposition of H$_{2}$S into H$_{3}$S
\cite{ScientificReports.4, arXiv.1509.03156E, PhysRevLett.114.157004}.

Theoretical predictions of high-temperature superconductivity rely on
knowledge of the stable structures of the system.  Global optimization
structure searching methods \cite{JCP.140.040901} can identify
previously unknown ground-state structures such as sulfur hydrides at
high pressures \cite{JCP.140.174712, PhysRevLett.114.157004,
  PhysRevB.91.180502, ScientificReports.4}.  Calculations based on
density functional theory (DFT) allow an accurate energetic evaluation
of the structures found within the potential energy landscape and an
assessment of their potential for superconductivity.  The experimental
discovery of high-temperature superconducting sulfur hydrides
demonstrates the predictive power of DFT-based structure searching and
EPC calculations, and suggests that more superconductors could be
discovered using this approach.  The role of theory is especially
important since under such extreme conditions sample synthesis and
\textit{in situ} measurements of properties are usually challenging.
Recently, the sister systems of selenium and tellurium hydrides were
predicted to exhibit high-$T_{c}$ superconductivity in hydrogen-rich
compounds stabilized above 100 GPa \cite{arXiv.1502.02607Z,
  arXiv.1503.00396Z}.  These proposals await experimental testing.

Here we explore the thermodynamic stability of pnictogen hydrides of
P, As and Sb at high pressures and their superconducting properties.
This investigation is in part motivated by the fact that the strength
of the covalent H--P bond (with a bond energy of 322 kJ/mol) is rather
similar to that of the H--S bond (bond energy of 363 kJ/mol) in the
molecular gas phase.  It is therefore reasonable to conjecture, within
the scenario that strong covalent bonding favors superconductivity
\cite{PhysRevB.91.060511, PhysRevB.92.060508}, that phosphorus
hydrides containing strong covalent bonds to H atoms and becoming
metallic under compression might exhibit similarly strong EPC and
potentially high-$T_{c}$ superconductivity.  To the best of our
knowledge, solid pnictogen hydrides, even at ambient conditions, have
only rarely been studied \cite{GazzettaChimicaItaliana.60.981,
  JPC.98.4973, ChemicalReviews.93.1623, InorganicChemistry}, although
the corresponding molecular gases (\textit{e.g.}, PH$_{3}$,
P$_{2}$H$_{4}$, AsH$_{3}$, and SbH$_{3}$) are well known in chemistry.

We find that thermodynamic stability of the gas molecules (PH$_{3} >$
AsH$_{3} >$ SbH$_{3}$) is reversed at high pressures, resulting in
decomposition enthalpies that decrease from P, As to Sb hydrides.
Except for the molecular solid phases stabilized by weak van der Waals
interactions at low pressures \cite{GazzettaChimicaItaliana.60.981,
  JPC.98.4973, ChemicalReviews.93.1623, InorganicChemistry}, P
hydrides are found to exhibit thermodynamic instability to
decomposition into the elements under pressures up to 400 GPa. For As
hydrides, stable compounds emerge only above 300 GPa.  Sb hydrides are
the most easily stabilized compounds above 150 GPa, which is readily
achievable in diamond anvil cells.  We have identified two stable
hydrogen-rich compounds, SbH$_{4}$ and AsH$_{8}$, as possible
high-temperature superconductors with a $T_{c}$ of 102 K at 150 GPa
and 141 K at 350 GPa, respectively.  Based on analysis of current and
prior theoretical work on other binary hydrides, we have established a
connection between ambient-pressure chemical properties and energetic
stability, structural features, superconducting properties, $etc.$\ of
hydrides upon pressures.  The trends implied by this connection may
help in discovering new binary and multinary high-pressure
hydrogen-rich superconductors, since it does not require knowledge of
the very different chemistries of elements at high pressures.  Using
trends in behavior to help in discovering superconductors has long
been recognized, starting with the rules set out by Matthias
\cite{matthias1957progress, PhysRev.109.280}.  Finally, we mention the
puzzling fact that although there have been more than a dozen
predictions of high-temperature superconductivity in compressed
hydrogen-rich compounds \cite{PNAS.109.6463, PNAS.107.15708,
  PhysRevLett.101.107002, PNAS.107.1317, JCP.140.174712,
  ScientificReports.4, arXiv.1501.06336F, arXiv.1502.02607Z,
  arXiv.1503.00396Z, arXiv.1503.08587L, arXiv.1507.02616S,
  arXiv.1504.01196D, JChemPhys.140.124707, PhysRevLett.110.165504,
  RSCAdv.5.45812}, sulfur hydride is the only system to date that has
been demonstrated to exhibit a higher superconducting temperature than
the record of 164 K set by cuprates \cite{PhysRevB.50.4260}.  The
second higher-$T_{c}$ material identified in experiments is silicon
hydride with $T_{c}$ = 17 K around 100 GPa, \cite{Science.319.1506},
although the origin of the superconductivity is still under debate
(likely from platinum hydride) \cite{Degtyareva2009, Dszc2015}.  The
chemical and physical properties of sulfur hydride that make it the
highest-$T_{c}$ superconductor found so far are not yet fully
understood.  Here we provide insight into this puzzle based on trends
derived from data for a wide range of high-pressure hydrides.

\section{\textbf{II. COMPUTATIONAL APPROACHES}}

Our investigation consists of determination of stable stoichiometries
and crystalline structures, and calculation of their superconducting
and related properties.  The hitherto unknown high-pressure phase
diagrams (at 0 K) of pnictogen (X) hydrides were explored by
performing very extensive structure searches for a set of
stoichiometries X$_{n}$H$_{m}$, and the most stable structure and
low-lying enthalpy metastable structures were identified for each
X$_{n}$H$_{m}$.  These calculations were performed using two leading
codes for first-principles structure prediction, CALYPSO
\cite{PhysRevB.82.094116, CPC.183.2063} and AIRSS
\cite{PhysRevLett.97.045504, JPhysCondensMatter.23.053201}.  The
energetically stable stoichiometries and the most stable structure at
each stoichiometry obtained from the two searching methods are in
agreement.  Over 100 stoichiometries were searched and a total of
about 100,000 structures were relaxed by minimizing the total energy 
within the P/H, As/H and Sb/H systems.

DFT calculations were performed using the VASP code
\cite{PhysRevB.54.11169} with projected-augmented-wave (PAW)
\cite{PhysRevB.50.17953} potentials, and the CASTEP code with
ultrasoft pseudopotentials.  The 1$s$ (H), 3$s$ and 3$p$ (P), 4$s$ and
4$p$ (As), and 5$s$ and 5$p$ (Sb) electrons were treated explicitly as
valence electrons.  We used the generalized gradient approximation
exchange-correlation functional of Perdew, Burke and Ernzerhof
\cite{PhysRevB.46.6671}.
Medium quality computational parameters were used when evaluating the
enthalpies of structures during the searches.  The low-lying enthalpy
structures were then further optimized using the PAW method and more
accurate computational parameters consisting of kinetic energy cutoff
energies of 650 eV (P/H and As/H) and 510 eV (Sb/H), and k-point
meshes of spacing 2$\pi \times$ 0.03 \AA$^{-1}$ or less. We checked
the energy convergence with respect to these parameters and found
convergence of the enthalpy differences to better than 1 meV/atom
level.

We calculated the phonon dispersion relations and EPC using linear
response theory as implemented in the Quantum ESPRESSO package
\cite{JPhysCondensMatter.21.395502}.  These calculations were
performed using norm-conserving pseudopotentials and a kinetic energy
cutoff of 100 Ry.  
For reliable evaluation of the double-delta function
integrals in the EPC calculations, dense k-point meshes with a spacing
of about 2$\pi \times$ 0.015 \AA$^{-1}$ were used, in combination with
the Methfessel-Paxton broadening scheme \cite{PhysRevB.40.3616} with a
broadening parameter of 0.05 Ry.  For the Brillouin sampling of phonon
momentum, a $q$-point mesh with a grid spacing of $\sim$2$\pi \times$
0.06 \AA$^{-1}$ was adopted.

\begin{figure}[ht]
\centerline{\includegraphics[width=3.5in]{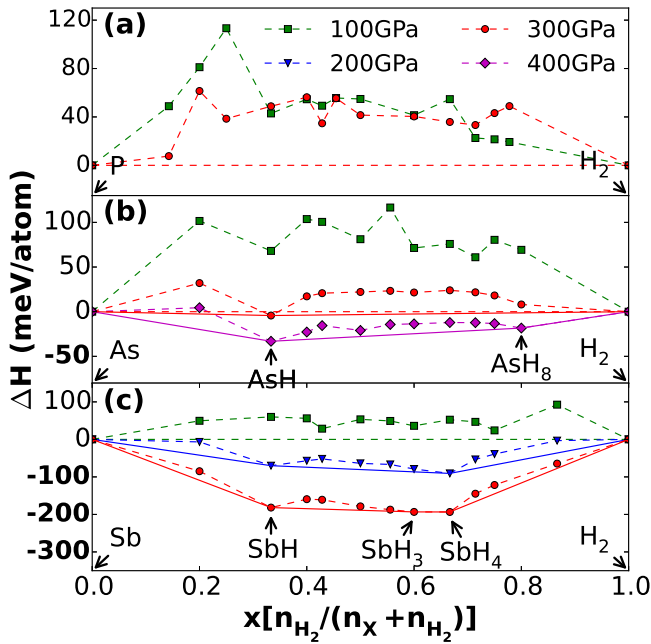}}
\caption{(color online) Calculated formation enthalpies $\Delta$H (in
  meV/atom) of various pnictogen hydrides with respect to
  decomposition into the elemental solids at 100 (squares), 200
  (triangles), 300 (circles) and 400 (diamond) GPa, respectively. The
  data for each stoichiometry corresponds to the lowest enthalpy
  structures from the searches. The structures of phases VI ($Pm\bar
  3m$) and VIII ($Im\bar 3m$) of P \cite{PhysRevB.77.024109}, IV
  ($Im\bar 3m$) of As \cite{PhysRevB.77.024109}, III ($Im\bar 3m$) of
  Sb \cite{PhysRevB.77.024109}, and the $P6_{3}m$ and $C2/c$
  structures of solid H$_{2}$ \cite{NaturePhysics.3.473} were used for
  evaluating $\Delta$H.}
\label{convexhull}
\end{figure}

\section{\textbf{III. RESULTS AND DISCUSSION}}

\textbf{3.1 Phase stabilities and stable crystalline structures of
  pnictogen hydrides at high pressures.}  Since H-rich compounds are
potentially more favorable for high-temperature superconductivity
\cite{PhysRevLett.92.187002}, we focused our CALYPSO searches on the
X$_{n}$H$_{m}$ ($n$=1$\sim$3 and $m$=1$\sim$8) stoichiometries with
higher H contents ($m \geq n$) containing up to 4 formula units per
simulation cell.  More diverse stoichiometries (with $m$ up to 13)
were considered in the AIRSS searches.  The main results from these
cross-checking search calculations are depicted in the hull diagrams
of Fig.\ \ref{convexhull}, which each contain data for pressures of
100--300 GPa.  For As hydrides (Fig.\ \ref{convexhull}b), searches at
the higher pressure of 400 GPa were performed to demonstrate the
stability of the AsH and AsH$_{8}$ structures above 300 GPa.
Only the lowest enthalpy structure at each stoichiometry studied is
shown.

We find a surprising trend in that at high pressures the thermodynamic
stability of pnictogen hydrides increases with the atomic number of
the pnictogen.  The P hydrides (Fig.\ \ref{convexhull}a) are found to
be unstable against decomposition into the elements, and pressure does
not have a noticeable effect on their stability.  The predicted
lowest-energy structures of several metastable stoichiometries are shown in
Supplementary Fig.\ S1 and Table S1.  As and Sb hydrides (Fig.\ \ref{convexhull}b
and \ref{convexhull}c) show a clear tendency to be stabilized by
increasing pressure; for As hydrides, a single stoichiometry stable
against elemental decomposition into the elements emerges at 300 GPa,
while for Sb hydrides, all the stoichiometries become stable above 200
GPa.  These results indicate that the relative stabilities of the
pnictogen hydrides obtained from the known chemical bonding strengths
at ambient conditions (P $>$ As $>$ Sb) are opposite to those found at
high pressures (P $<$ As $<$ Sb).  We find two stable binary As
hydrides on the convex hull at 400 GPa, AsH and AsH$_{8}$.  We find
three compounds on the convex hull for Sb hydrides at 300 GPa and two
at 200 GPa.  These are SbH, SbH$_{3}$ (at 300 GPa only) and SbH$_{4}$.
Note that the much lower formation enthalpies of Sb hydrides
($\sim$200 meV/atom more stable) demonstrates their high stability
under compression.  We evaluated the effect of zero point energy (ZPE)
on the phase stability by calculating harmonic phonon spectra for Sb
hydrides, as shown in Supplementary Fig.\ S2.  Inclusion of the ZPE
gives rise to a negligible change in the convex hulls.

The structures of the predicted stable compounds are shown in Fig.\
\ref{stableStructure}a-e (see Supplementary Table S2 for detailed
structural information), and the pressure ranges within which they are
stable are depicted in Fig.\ \ref{stableStructure}f.  The stable
structures of the compounds remain unchanged throughout their stable
pressure ranges.  The AsH hydride (Fig.\ \ref{stableStructure}a)
adopts a structure of $Cmcm$ symmetry which is stable above 300
GPa. It consists of a three-dimensional As-H network in which both As
and H atoms are approximately five-fold coordinated. The lowest
enthalpy structure of AsH$_{8}$ (Fig.\ \ref{stableStructure}b) of
$C2/c$ symmetry is stabilized above 350 GPa.  It is formed from the
primary motifs of irregular AsH$_{16}$ polyhedra, connected with each
other in a three-dimensional network.  Each corner H atom of the
AsH$_{16}$ polyhedron is linked to another H from the adjacent
polyhedron. The short contact between two H atoms leads to formation
of an intriguing quasi-molecular H$_{2}$-unit with a bond length of
$\sim$0.8--0.9 \AA. The formation of such H$_{2}$-units in compressed
hydrides is also seen in systems with Si, Ge, Sn, Li, Ca, Te
\cite{PNAS.106.17640, PNAS.109.6463, PhysRevLett.101.107002,
  PNAS.107.1317, PhysRevLett.98.117004, arXiv.1503.00396Z}, etc.

\begin{figure}[ht]
\centerline{\includegraphics[width=3.5in]{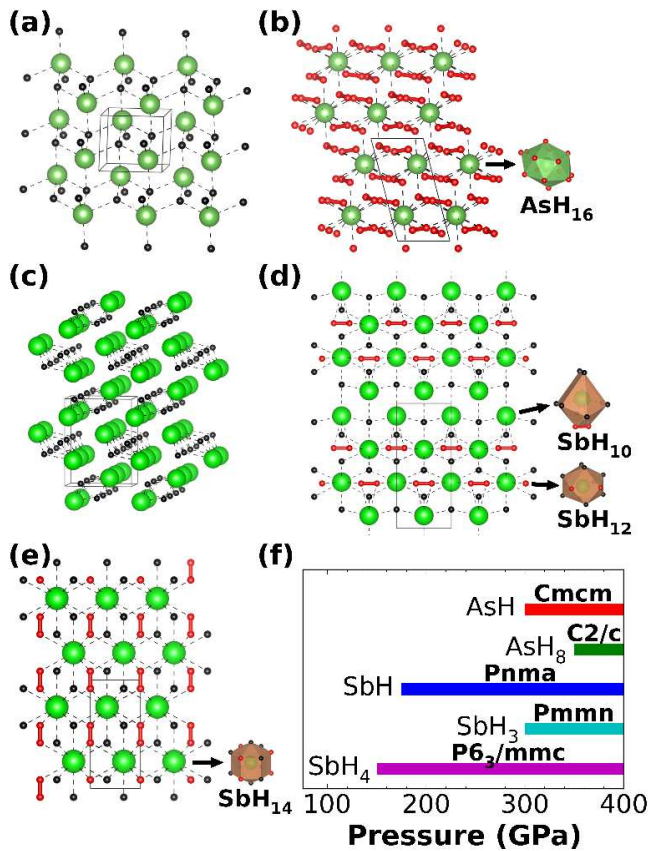}}
\caption{(color online) Structures of the predicted stable pnictogen
  hydrides: (a) AsH (space group $Cmcm$), (b) AsH$_{8}$ ($C2/c$), (c)
  SbH ($Pnma$), (d) SbH$_{3}$ ($Pmmn$); (e) SbH$_{4}$
  ($P6_{3}/mmc$). The As/Sb--H bonds are depicted by dashed lines and
  H--H covalent bonds as red sticks. The corresponding stable pressure
  range of each compound is shown in (f).}
\label{stableStructure}
\end{figure}

The energetically favored structure of SbH (Fig.\
\ref{stableStructure}c, stable above 175 GPa) has space group $Pnma$
and is composed of chain-like Sb-H motifs with each Sb coordinated by
three H atoms.  The H-rich compounds SbH$_{3}$ (Fig.\
\ref{stableStructure}d) and SbH$_{4}$ (Fig.\ \ref{stableStructure}e)
have $Pmmn$ and $P6_{3}/mmc$ symmetries, respectively.  While the
former is marginally stable (very close to the convex hull formed by
SbH and SbH$_{3}$ in Fig.\ \ref{convexhull}c) at 300 GPa, the latter
is robustly stable above 150 GPa. SbH$_{3}$ is composed of irregular
polyhedral SbH$_{10}$ and SbH$_{12}$ motifs which form its
three-dimensional topology.  Bridging the adjacent SbH$_{12}$ motifs
in a two-dimensional fashion gives rise to quasi-molecular
H$_{2}$-units (in red) with a rather long bond length of $\sim$0.91
\AA, compared with the molecular H$_2$ bond length of 0.74 \AA.
SbH$_{4}$ is a highly symmetrical hexagonal compound that consists of
SbH$_{14}$ octadecahedra connected via shared H atoms at the corners
to form a three-dimensional network.  There are two types of H atoms
occupying the $4e$ (black) and $4f$ (red) Wyckoff sites, respectively.
Each $4e$ H atom is coordinated by three Sb atoms, and the $4f$ H
atom, while coordinated by two Sb atoms, has a close contact with
another $4f$ H atom, forming quasi-molecular H$_{2}$-units with a bond
length of $\sim$0.83 \AA.  SbH$_{4}$ resembles the TeH$_{4}$ compound
with the same stoichiometry which adopts the $R\bar{3}m$ phase
\cite{arXiv.1503.00396Z}.

A Bader charge density analysis \cite{JPhysCondensMatter.21.084204}
shows substantial charge transfer from As/Sb to H (see Supplementary
Table S3).  This indicates that the As/Sb--H bonds have substantial
ionic character.  On the other hand, plots of the electron
localization function (ELF) (see Supplementary Fig.\ S5) show
material-dependent (see below) charge concentration between As/Sb and
H atoms.  Hence the predicted stable compounds are dominated by mixed
covalent and ionic interactions.

\begin{figure}[ht]
\centerline{\includegraphics[width=3.5in]{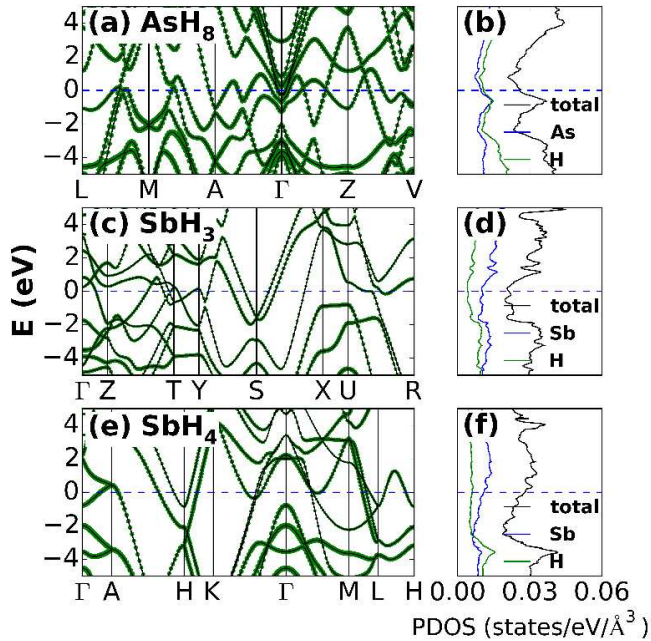}}
\caption{(color online) Calculated band structures and projected
  densities of states of stable H-rich pnictogen hydrides: (a,b)
  AsH$_{8}$ at 350 GPa, (c,d) SbH$_{3}$ at 300 GPa and (e,f) SbH$_{4}$
  at 150 GPa.  The green circles in the band structures represent the
  orbital projections of the electronic states onto H atoms.}
\label{stableBandStructure}
\end{figure}

\textbf{3.2 Electronic structure, phonons, electron-phonon coupling
  and superconductivity of identified stable pnictogen hydrides.}  The
DFT calculations indicate that all of the compounds are metallic
within their stable pressure ranges.  Fig.\ \ref{stableBandStructure}
shows band structures and projected densities of states (DOS) for
three H-rich compounds, AsH$_{8}$, SbH$_{3}$ and SbH$_{4}$ (see
Supplementary Fig.\ S3 for results for H-poor AsH and SbH).  The Fermi
levels (E$_{f}$) of AsH$_{8}$ (Fig.\ \ref{stableBandStructure}a-b) and
SbH$_{4}$ (Fig.\ \ref{stableBandStructure}e-f) are located at
shoulders of a peak in the DOS, leading to a rather large value of the
DOS at E$_{f}$ (N(E$_{f}$)).  Moreover, substantial H-derived states
exist in proximity to E$_{f}$, which mimics solid metallic hydrogen.
These features are potentially favorable for high-temperature
superconductivity.  This is, however, not the case for SbH$_{3}$
(Fig.\ \ref{stableBandStructure}c-d) in which E$_{f}$ lies near the
bottom of a broad valley in the DOS, which gives a low N(E$_{f}$), and
only a small amount of H derived states at E$_{f}$.

\begin{figure}[ht]
\centerline{\includegraphics[width=3.5in]{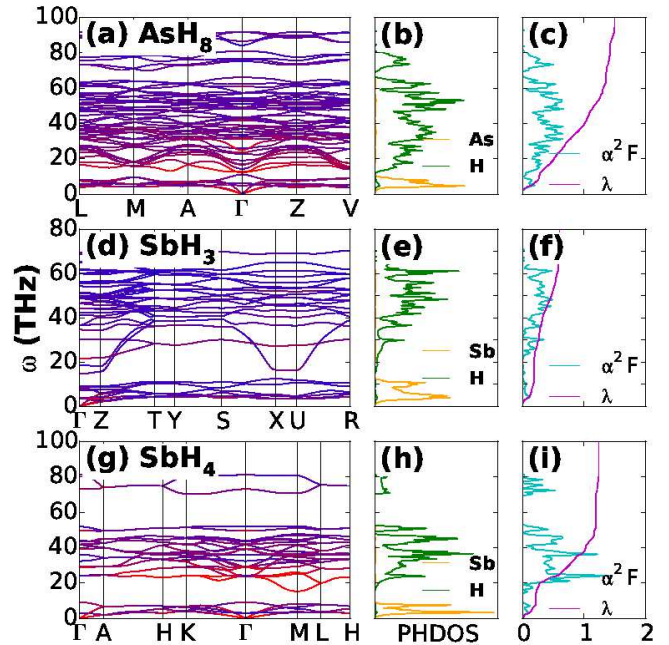}}
\caption{(color online) Calculated phonon dispersion curves (a,d,g),
  projected phonon density of states (PHDOS) (b,e,h), Eliashberg EPC
  spectral functions $\alpha^{2}$F$(\omega)$ and its integral
  $\lambda(\omega)$ (c,f,i) of AsH$_{8}$ at 350 GPa, SbH$_{3}$ at 300
  GPa and SbH$_{4}$ at 150 GPa, respectively. The proportion of red
  color in the phonon dispersion curves represents the magnitude of
  the EPC parameter $\lambda_{n,q}(\omega)$ at each phonon mode
  ($n,q$).}
\label{stableDispersion}
\end{figure}

As shown in the phonon spectra of Fig.\ \ref{stableDispersion} (a:
AsH$_{8}$, d: SbH$_{3}$ and g: SbH$_{4}$) and Supplementary Fig.\ S4
(AsH and SbH), all of the compounds are dynamically stable as
demonstrated by the absence of imaginary phonon frequencies.  The
projected phonon DOS of SbH$_{4}$ (Fig.\ \ref{stableDispersion}h)
shows three well separated regions: the low-frequency vibrations
associated with heavy Sb atoms (below 15 THz), higher frequency
wagging, bending and stretching modes derived from the H atom bonded
to Sb (between 15 and 65 THz) and high-frequency intra-molecular H--H
stretching modes from the quasi-molecular H$_{2}$-units (around 80
THz).  The H--H stretching modes of SbH$_{3}$ (Fig.\
\ref{stableDispersion}e) are lower in frequency and they merge into
the medium-frequency region because of the larger bond lengths of the
H$_{2}$-units.  In AsH$_{8}$ (Fig.\ \ref{stableDispersion}b), two
groups of H--H stretching modes appear, which correspond to two types
of H$_{2}$ unit with different bond lengths.

We performed EPC calculations for the predicted stable compounds to
probe their superconducting properties.  Fig.\ \ref{stableDispersion}
shows the EPC parameters for each phonon mode (a,d,g), the Eliashberg
EPC spectral function $\alpha^{2}$F$(\omega)$ and its integral
$\lambda(\omega)$ (c,f,i) for AsH$_{8}$, SbH$_{3}$ and SbH$_{4}$
(results for AsH and SbH are shown in Supplementary Fig.\ S4).  Fig.\
\ref{Tc-lambda-NEf-P}a summarizes data for the total EPC parameters
$\lambda$ and their pressure dependence.  AsH$_{8}$ and SbH$_{4}$ are
H-rich compounds with rather high $\lambda$ values above 1.0, while
SbH$_{3}$ has a lower value of about 0.5.  For all three H-rich
compounds the intermediate-frequency H-derived wagging, bending and
stretching phonons and low-frequency vibrations from the heavy As/Sb
atoms, rather than the high-frequency phonons derived from the
H$_{2}$-units, dominate the contributions to the total EPC.

Within the framework of strong-coupling BCS theory
\cite{PhysRev.167.331}, the total EPC parameter is proportional to the
product of N(E$_{f}$) and the squared electronic matrix element $<$
I$^{2}>$ averaged over the Fermi surface.  The calculated values of
N(E$_{f}$) for each compound are shown in Fig.\
\ref{Tc-lambda-NEf-P}b. While AsH$_{8}$ and SbH$_{4}$ have rather high
values of N(E$_{f}$), SbH$_{3}$ exhibits quite a low N(E$_{f}$).
These results accord with their significantly different $\lambda$
values. The high N(E$_{f}$) in H-poor AsH arises predominantly from
As-induced states (Supplementary Fig.\ S3b) and results in a small
value of $\lambda$ similar to that of superconducting solid As under
compression \cite{PhysRevB.46.5523, JETPL.10.55B}.  Besides the
difference in N(E$_{f}$), we find that the strengths of the covalent
components of As/Sb--H bonds (with mixed covalent and ionic nature)
vary considerably among the compounds, as indicated by the ELF data
shown in Supplementary Fig.\ S5.  The Sb--H bonds in SbH$_{3}$ are
much weaker than those of SbH$_{4}$ and the As-H bonds in
AsH$_{8}$. The weaker Sb--H bonds in SbH$_{3}$ in principle correspond
to a low deformation potential, \textit{i.e.}, a smaller modification
of the electronic structure, with respect to the motion of H
atoms. This leads to a relatively small $<$ I$^{2}>$, and thus a low
$\lambda$.

\begin{figure}[ht]
\centerline{\includegraphics[width=3.5in]{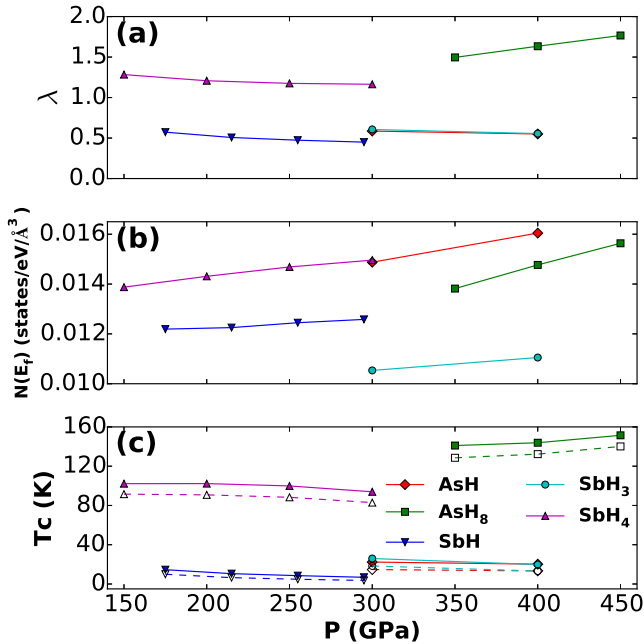}}
\caption{(color online) The total EPC parameter $\lambda$ (a),
  electronic DOS at the Fermi level N(E$_{f}$) (b) and superconducting
  critical temperature $T_{c}$ (c) as a function of pressure for the
  predicted stable compounds. For the $T_{c}$ calculations, two
  different screened Coulomb repulsion parameters of $\mu^{*}$ = 0.1
  (filled symbols and solid lines) and 0.13 (open symbols and dashed
  lines) are used for comparison.}
\label{Tc-lambda-NEf-P}
\end{figure}

The superconducting $T_{c}$ values of the stable compounds are
evaluated using the Allen-Dynes modified McMillan equation
\cite{PhysRevB.12.905} with calculated logarithmic average frequency
($\omega_{log}$) and a typical choice of screened Coulomb repulsion
parameter of $\mu^{*} = 0.1$. The results shown in Fig.\
\ref{Tc-lambda-NEf-P}c (see Supplementary Table S4 for numerical
values) are consistent with the trends in $\lambda$ (Fig.\
\ref{Tc-lambda-NEf-P}a).  AsH$_{8}$ and SbH$_{4}$ exhibit quite high
$T_{c}$ values of around 150 K and 100 K, respectively.  The other
compounds have moderate $T_{c}$ values of $\sim$20 K or below.  The
value of $T_{c}$ in SbH$_{4}$ decreases slightly with pressure, while
AsH$_{8}$ shows an increase.  Using a standard value of $\mu^{*} =
0.13$ for metallic hydrides \cite{PhysRevLett.92.187002}, we find a
small decrease in $T_{c}$ with pressure, $\sim$8--11$\%$ in the
higher-$T_{c}$ compounds of AsH$_{8}$ and SbH$_{4}$.

\begin{figure}[ht]
\centerline{\includegraphics[width=3.5in]{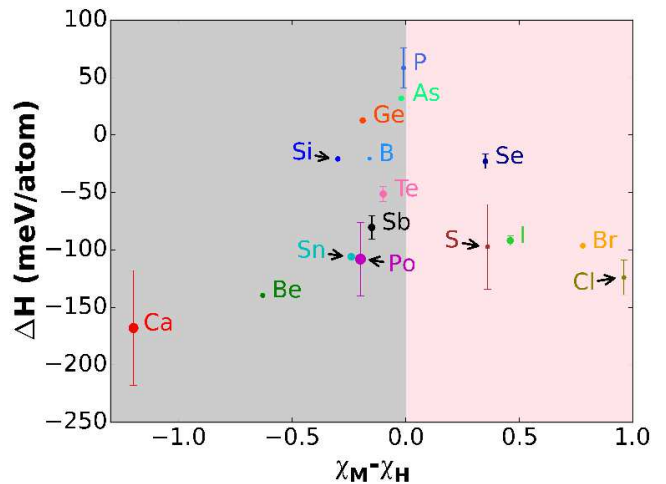}}
\caption{(color online) Formation enthalpies $\Delta$H (in meV/atom)
  of main-group-element (M) binary hydrides \cite{PNAS.109.6463,
    PNAS.107.15708, PhysRevLett.101.107002, PNAS.107.1317,
    arXiv.1502.02607Z, arXiv.1503.00396Z, arXiv.1503.08587L,
    arXiv.1507.02616S, ChemSci.6.522, arXiv.1504.01196D,
    JChemPhys.140.124707, PhysRevLett.110.165504,
    PhysRevLett.114.157004} at 200 GPa with respect to decomposition
  into the constituent elemental solids, plotted against the
  electronegativity difference between M and H at ambient pressure,
  $\chi_{M} - \chi_{H}$. The electronegativity values are taken from
  Pauling's scale \cite{lide2004crc}.  When several stable phases with
  different stoichiometries exist, the averaged formation enthalpy is
  shown, and the endpoints of the error bars represent the minimum and
  maximum values of $\Delta$H.  The sizes of the circles are
  proportional to the empirical atomic radii \cite{JChemPhys.41.3199}.
  The regions corresponding to positive and negative $\chi_{M} -
  \chi_{H}$ are shown in the background colors of pink and gray,
  respectively.}
\label{electroNegativityForamtionEnthalpy}
\end{figure}

\textbf{3.3 Deduced general chemical trends of high-pressure
  hydrogen-containing superconductors.}  Gathering data for pnictogen
hydrides and other hydrides \cite{PNAS.109.6463, PNAS.107.15708,
  PhysRevLett.101.107002, PNAS.107.1317, arXiv.1502.02607Z,
  arXiv.1503.00396Z, arXiv.1503.08587L, arXiv.1507.02616S,
  ChemSci.6.522, arXiv.1504.01196D, PhysChemChemPhys.17.19379,
  JChemPhys.140.124707, PhysRevLett.110.165504,
  PhysRevLett.114.157004} allows us to deduce useful physical
principles about the phase stability and occurrence of
high-temperature superconductivity in H-containing materials under
pressure.  For clarity, here we consider only the binary hydrides
M$_{n}$H$_{m}$ formed by main-group elements M with $sp$ bonding.  We
find that the Pauling electronegativity difference between M and H at
ambient pressure, $\chi_{M} - \chi_{H}$, is a good ``descriptor'' of
high-pressure phase stability, structural features and superconducting
properties of the hydrides.  Specifically, we find that the following
properties of M$_{n}$H$_{m}$ are related to $\chi_{M} - \chi_{H}$:

(i) \textit{Energetic stability against elemental decomposition}.
Fig.\ \ref{electroNegativityForamtionEnthalpy} shows formation
enthalpies of various hydrides at 200 GPa as a function of $\chi_{M} -
\chi_{H}$.  For cases in which more than one stable stoichiometry
exists, the averaged formation enthalpy (among all the
stoichiometries) is shown, accompanied by an error bar indicating the
maximum and minimum enthalpy.  The diameter of the circular symbol is
proportional to the atomic radius of M \cite{JChemPhys.41.3199}.  This
divides hydrides into two classes with negative and positive $\chi_{M}
- \chi_{H}$, respectively.  Considering the extreme cases, H$_{2}$O
with $\chi_{M} - \chi_{H} \gg 0$ has strong covalent O--H bonds,
whereas LiH with $\chi_{M} - \chi_{H} \ll 0$ has strong ionic Li--H
bonding.  It may thus be reasonable to conjecture that the hydrides
with $\chi_{M} - \chi_{H} > 0$ prefer to form covalent M--H bonds,
while those with $\chi_{M} - \chi_{H} < 0$ are more ionic in nature.
In the regions of both $\chi_{M} - \chi_{H} > 0$ and $\chi_{M} -
\chi_{H} < 0$, we observe a general trend that the formation enthalpy
decreases (\textit{i.e.}, stability increases) with $|\chi_{M} -
\chi_{H}|$.  For $\chi_{M} - \chi_{H} > 0$, this can be explained by
the larger electronegativity of M which corresponds to strongly
covalent M--H bonds, and therefore to higher stability of the
hydrides.  In this region, the atomic radius can also affect the
covalent bond strength, and therefore have a critical role in
determining the stability.  For instance, despite the similar
electronegativities of S and Se, S hydrides have much lower formation
enthalpies than Se hydrides (by about 50$\sim$150 meV/atom). We
attribute this to the smaller atomic radius of S favoring stronger
covalent H--S bonds.  In the region $\chi_{M} - \chi_{H} < 0$, the
smaller electronegativity of M and larger value of $|\chi_{M} -
\chi_{H}|$ leads to substantial charge transfer from the cationic M to
anionic H.  On the one hand, this increases the Madelung energy of the
ionic component of the M--H bonding, which increases the stability.
On the other hand, the charge transfer is favorable for the formation
of quasi-molecular H units in the interstitial regions (see below).  This
lowers the enthalpy of hydrides and make them more stable.  This
explanation accords with the increased thermodynamic stability of P,
As, and Sb hydrides, as demonstrated above.

\begin{figure}[ht]
\centerline{\includegraphics[width=3.5in]{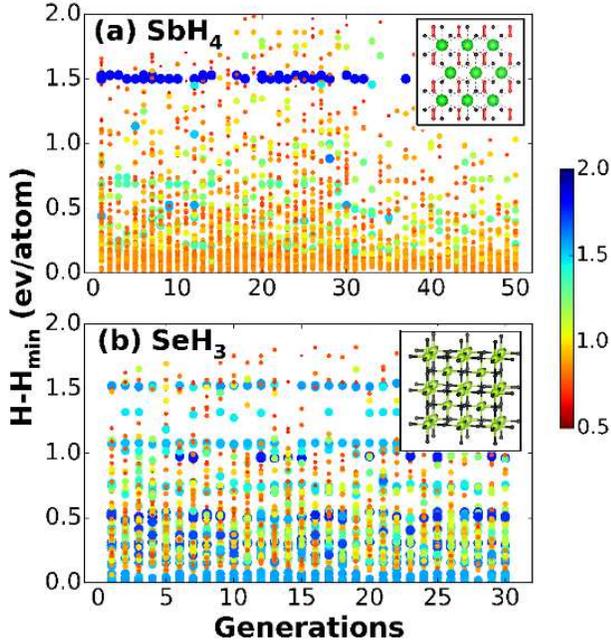}}
\caption{(color online) The structures generated are sorted according
  to enthalpy during the CALYPSO structure searches for SbH$_{4}$ (a)
  and SeH$_{3}$ (b) at 200 GPa, as a function of search
  generations. The color-coding represents the minimum distance
  between two H atoms, which is also proportional to the size of the
  circle. The insets show the predicted ground-state structures of
  SbH$_{4}$ and SeH$_{3}$.}
\label{evolution}
\end{figure}

(ii) \textit{Key structural features}.  For the hydrides with
$\chi_{M} - \chi_{H} < 0$, for example our predicted AsH$_{8}$,
SbH$_{3}$ and SbH$_{4}$ phases, there is substantial charge transfer
from the relatively electropositive M to the electronegative H
(Supplementary Table S3).  The M--H bonds have mixed covalent and
ionic nature.  With such weaker M--H bonding, the H atoms are prone to
move to the interstitial regions under pressures.  Therefore,
substantial energy gain may be achieved via the formation of H--H
covalent bonds in the interstitial regions.  This is especially the
case under pressure where large non-bonded anions are
unfavorable. This leads to the formation of intriguing quasi-molecular
H-units in H-rich compounds \cite{arXiv.1503.00396Z, PNAS.107.1317,
  PhysRevLett.101.107002, PNAS.107.15708, PNAS.109.6463,
  PNAS.106.17640, PhysRevLett.106.237002}. Turning to $\chi_{M} -
\chi_{H} > 0$, for instance S \cite{JCP.140.174712,
  ScientificReports.4, PhysRevLett.114.157004} and Se
\cite{arXiv.1501.06336F, arXiv.1502.02607Z} hydrides, the strong polar
covalent M--H bonds dominate.  Under such conditions, all cationic H
atoms are tightly bonded by the M atoms with large electronegativity,
and thus the formation of H--H bonds is energetically unfavourable.
The quasi-molecular H-unit is absent in the hydrides.  The distinction
between this key structural behavior in the two regions is reflected
in Fig.\ \ref{evolution}, which shows the evolution of the CALYPSO
search for SbH$_{4} (\chi_{M} - \chi_{H} < 0)$ and SeH$_{3} (\chi_{M}
- \chi_{H} > 0)$ at 200 GPa.  Usually tens of generations are required
to determine the global minimum in the CALYPSO calculations.  The
structures at each generation are sorted according to their enthalpy.
The color coding represents the minimum distance between two H atoms
in each structure.  All of the low enthalpy structures of SbH$_{4}$
contain quasi-molecular H$_{2}$ units with short contacts
(0.8$\sim$0.9 \AA) between neighboring H atoms.  In contrast, in
SeH$_{3}$ the low-lying structures exhibit large minimum H--H
distances (over $1.5$ \AA) as three-dimensional covalent H-Se networks
are formed.

\begin{figure}[ht]
\centerline{\includegraphics[width=3.5in]{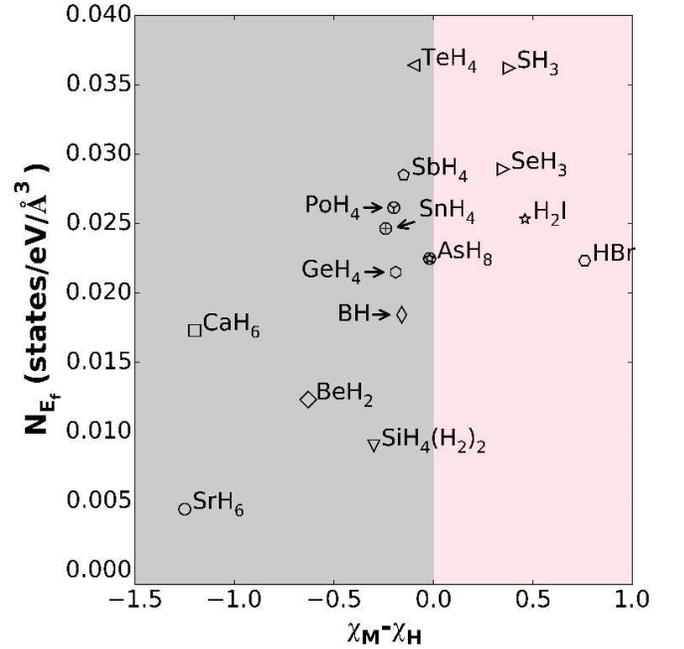}}
\caption{(color online) Calculated electronic DOS at the Fermi level
  N(E$_{f}$) of main-group-element binary hydrides
  \cite{PNAS.109.6463, PNAS.107.15708, PhysRevLett.101.107002,
    PNAS.107.1317, arXiv.1502.02607Z, arXiv.1503.00396Z,
    arXiv.1503.08587L, arXiv.1507.02616S, ChemSci.6.522,
    PhysChemChemPhys.17.19379, JChemPhys.140.124707,
    PhysRevLett.110.165504, RSCAdv.5.45812, PhysRevLett.114.157004} at
  200 GPa, as a function of the electronegativity difference between M
  and H at ambient pressure, $\chi_{M} - \chi_{H}$.  When several
  stable phases with different stoichiometries exist, we choose the
  one with the highest predicted superconducting $T_{c}$. Similar to
  Fig.\ \ref{electroNegativityForamtionEnthalpy}, the regions of
  $\chi_{M} - \chi_{H} > 0$ and $\chi_{M} - \chi_{H} < 0$ are in pink
  and grey, respectively.}
\label{StatesEfermi}
\end{figure}

(iii) \textit{Density of states at the Fermi level}.  As demonstrated
in Fig.\ \ref{Tc-lambda-NEf-P}b, N(E$_{f}$) is a critical parameter in
determining the strength of the EPC.  Fig.\ \ref{StatesEfermi} shows
the values of N(E$_{f}$) for the binary hydrides that are predicted to
be superconductors \cite{PNAS.109.6463, PNAS.107.15708,
  PhysRevLett.101.107002, PNAS.107.1317, arXiv.1502.02607Z,
  arXiv.1503.00396Z, arXiv.1503.08587L, arXiv.1507.02616S,
  ChemSci.6.522, PhysChemChemPhys.17.19379, JChemPhys.140.124707,
  PhysRevLett.110.165504, RSCAdv.5.45812, PhysRevLett.114.157004}, as
a function of $\chi_{M} - \chi_{H}$ at 200 GPa.  To make a fair
comparison and avoid potential errors arising from different
theoretical calculations, we take structural data from Ref.\
\onlinecite{PNAS.109.6463, PNAS.107.15708, PhysRevLett.101.107002,
  PNAS.107.1317, arXiv.1502.02607Z, arXiv.1503.00396Z,
  arXiv.1503.08587L, arXiv.1507.02616S, ChemSci.6.522,
  PhysChemChemPhys.17.19379, JChemPhys.140.124707,
  PhysRevLett.110.165504, RSCAdv.5.45812, PhysRevLett.114.157004}, and
perform full structural optimizations and calculate N(E$_{f}$) using
the plane-wave PAW method.  In general, we find that high values of
N(E$_{f}$) appear in the region with relatively small magnitudes of
$\chi_{M} - \chi_{H}$.  The compounds with higher values of N(E$_{f}$)
in the two regions are TeH$_{4}$, SbH$_{4}$ $(\chi_{M} - \chi_{H} <
0)$ and SH$_{3}$, SeH$_{3}$ $(\chi_{M} - \chi_{H} > 0)$, respectively.
The highly symmetric structures of these hydrides (TeH$_{4}$:
$P6/mmm$; SbH$_{4}$: $P6_{3}/mmc$; S/SeH$_{3}$: $Im\bar3m$), which in
principle give rise to high degeneracy of the band structure at
special $k$-points, may be responsible for the high values of
N(E$_{f}$).  They are all predicted to be good superconductors with
values of $\lambda$ above 1 \cite{ScientificReports.4,
  arXiv.1503.00396Z, PhysRevLett.114.157004, arXiv.1502.02607Z,
  arXiv.1501.06336F}.

\begin{figure}[ht]
\centerline{\includegraphics[width=3.5in]{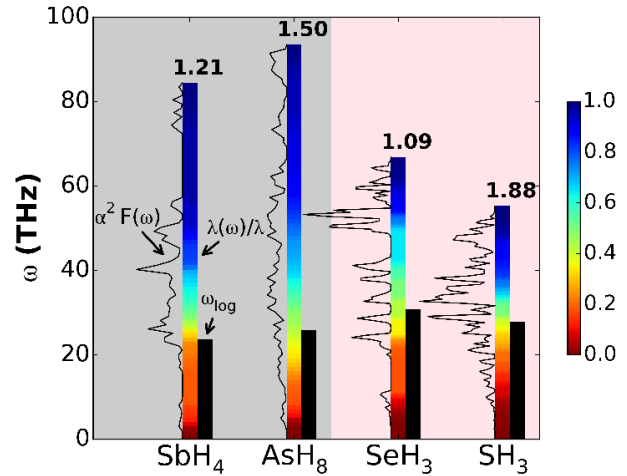}}
\caption{(color online) The Eliashberg EPC spectral function
  $\alpha^{2}$F$(\omega)$ (black lines), its integral
  $\lambda(\omega)$ (colorful scale), and logarithmic average
  frequency $\omega_{log}$ (black scale) for typical hydrides in the
  regions of $\chi_{M} - \chi_{H} > 0$ (SH$_{3}$ and SeH$_{3}$) and
  $\chi_{M} - \chi_{H} < 0$ (AsH$_{8}$ and SbH$_{4}$). For comparison
  among different compounds, the $\lambda(\omega)$ is normalized and divided by the total $\lambda$ (indicated for each compound).}
\label{epc}
\end{figure}

(iv) \textit{Features of the electron-phonon coupling}.  Fig.\
\ref{epc} shows the Eliashberg spectral function
$\alpha^{2}$F$(\omega)$ (black lines), its normalized integral $\lambda(\omega)$
(divided by the total $\lambda$, colorful scale), and
logarithmic average frequency $\omega_{log}$ (black scale) of typical
compounds from the two regions.  In the $\chi_{M} - \chi_{H} < 0$
region (AsH$_{8}$ and SbH$_{4}$), the total EPC originates mainly from
the medium-frequency phonons, predominantly wagging, bending, and
stretching modes derived from the H bonded to As/Sb.  The higher
frequency H--H stretching modes of the quasi-molecular H$_{2}$-units
make fairly small contributions to the EPC.  Therefore the integral
$\lambda(\omega)$ becomes saturated (in blue) at the upper-middle part
of the $\omega$ range.  Turning to the region of $\chi_{M} - \chi_{H}
> 0$ (SH$_{3}$ and SeH$_{3})$, the EPC arises mainly from the
high-frequency H-derived phonons of covalent M--H networks.  The
integral $\lambda(\omega)$ thus becomes saturated close to the upper
boundary of $\omega$.  These features of the EPC can be ascribed to
the different chemical bonding and structural features of the hydrides
in the two regions.  Note that $\omega_{log}$, the important quantity
determining the superconducting $T_{c}$, shows higher values in the
$\chi_{M} - \chi_{H} > 0$ region than in the $\chi_{M} - \chi_{H} < 0$
region, 
despite the evidently lower maximum $\omega$ of the former region than that of the latter.
This demonstrates that simply
hardening the phonons is not always effective in raising $T_{c}$.  The
reason originates from the distinct dependence of
$\alpha^{2}$F$(\omega)$ on $\omega$ in the two regions, since only the
phonons contributing to the EPC are counted when evaluating $T_{c}$.

\textbf{3.4 More relevant discussion.}  Since Ashcroft first proposed
metallic hydrides as potentially good superconductors at high
pressures \cite{PhysRevLett.92.187002}, high-$T_{c}$ (above 100 K)
superconductivity has been predicted in a number of high-pressure
H-rich compounds \cite{PNAS.109.6463, PNAS.107.15708,
  arXiv.1502.02607Z, arXiv.1503.00396Z, ScientificReports.4,
  PhysRevLett.114.157004}.  The S hydride is one of these predictions
\cite{JCP.140.174712, ScientificReports.4} that has been confirmed
experimentally \cite{Nature.525.73, arXiv.1509.03156E}.  In other
systems, either low temperature superconductivity
\cite{Science.319.1506, JPhysChemC.119.18007} or no superconductivity
\cite{PhysRevLett.100.045504, PhysicaBCondensedMatter.329.1312} was
found.  While the underlying reasons for this are not yet settled, one
possibility is that the predicted superconducting compounds did not
form in high-pressure experiments.  Confining hydrides or mixtures of
hydrogen and other substances in diamond anvil cells up to ultrahigh
pressure conditions is technically challenging and, due to the
technical limitations, experimental syntheses may fail even though the
theoretically predicted structures are energetically stable.  This is
highly likely for hydrides with small formation enthalpies, such as
those of Si, Ge, B, Se, \textit{etc.}, as indicated in Fig.\
\ref{electroNegativityForamtionEnthalpy}.

It is also possible that kinetic processes under pressure inhibit
formation of some of the predicted compounds.  If so, a highly
symmetric structure may be more likely to be synthesised because the
energetic barrier for the transition between the symmetric structure
and other isomers is usually large.  This suggests a strong kinetic
stability with respect to variations in the external conditions.  From
this point of view, S hydrides are indeed favorable for experimental
synthesis since the stable SH$_{3}$ structure has simultaneously a
large formation enthalpy ($\sim$135 meV/atom at 200 GPa) and high
symmetry (space group $Im\bar3m$). In this sense, the SbH$_{4}$ phase
predicted in the present work is quite promising for possible
experimental synthesis, in view of its large formation enthalpy
($\sim$90 meV/atom at 200 GPa) and highly symmetric structure (space
group $P6_{3}/mmc$).  
Certainly the larger barrier associated with the highly symmetric structure may make its synthesis challenging.
This can be basically overcome
by for instance sufficient thermal energy or catalytic assistant.

Turning to trends, results from the current and previous theoretical
studies reveal a connection between the thermodynamic stability,
structural features and superconducting properties of high-pressure
hydrides, and the Pauling electronegativity differences between H and
the other elements.  This is a remarkable and at first sight strange
finding, since high-pressure chemistry is well known to be radically
different from that at ambient pressure \cite{van2008high}.  Our
findings indicate that the ambient-pressure chemical properties
determine to a large extent the high-pressure behavior.

Electropositive elements (with $\chi_{M} - \chi_{H} \ll 0$ in our
classification), such as alkaline and alkaline earth metals, form
highly ionic ambient-pressure hydrides characterized by large charge
transfers and structures similar to fluorides. In particular, the
crystals are stabilized by the Madelung energy of the cationic M and
anionic H ions. The anionic H in these materials are well separated
from other H ions and the H vibrational frequencies are relatively
low.  At high pressures the distances between H anions are
significantly shortened, but the weak ionic M--H bonds allow H atoms
to move to the interstitial regions.  These lead to stabilization of
the H--H covalent bonds in the interstitial region.  Thus
quasi-molecular aggregation of H emerges \cite{PNAS.106.17640,
  PhysRevLett.106.237002, PNAS.109.6463, PhysChemChemPhys.17.19379}
and high H vibrational frequencies occur in the high-pressure phases.

Non-metallic strongly electronegative elements (with $\chi_{M} -
\chi_{H} \gg 0$) often form weakly bound insulating H-containing
molecular solids at ambient pressure, \textit{e.g.}, NH$_{3}$,
H$_{2}$O, H$_{2}$S, \textit{etc}. While strong covalent M--H bonds
dominate molecules, the intermolecular bonds characterizing cohesion
of molecular solids are weak van der Waals bonds and hydrogen bonds.
Under compression, with decreasing intermolecular distances and
accompanying hydrogen-bond symmetrization \cite{PhysRevLett.69.462,
  PhysRevB.54.15673, JChemPhys.84.2771}, solids with covalent M--H
bonding networks eventually become energetically favored.  Because of
the strong polar covalent M--H bonds, the H--H bonds and aggregation
of H atoms are energetically unfavourable.

As for the elements with moderate electronegativities comparable to
that of H ($\chi_{M} - \chi_{H} \sim 0$), the hydrides formed are
intermediate between the above two cases.  The M--H bonds have mixed
covalent and ionic characters, and the former is not strong enough to
stabilize covalently bonded M--H networks. In the hydrides with
$\chi_{M} - \chi_{H} < 0$ (Si, Ge, Sn, Te, Sb, $etc.$), the anionic
nature of H atoms (due to the charge transfer from M to H) and the
tendency of H atoms to move to interstitial regions (due to the
relatively weak M--H bonds) under pressure allows aggregation of H via
formation of H--H covalent bonds in H-rich phases.  This mechanism
helps to stabilize hydrides.  The stability increases with the
magnitude of $\chi_{M} - \chi_{H}$ (as in Fig.\
\ref{electroNegativityForamtionEnthalpy}), which determines the
Madelung energy of the ionic interaction and the capability of
hydrides to form aggregates of H.  For As and P hydrides with small
$|\chi_{M} - \chi_{H}|$, only limited charge transfer from As/P to H
can occur due to the small electronegativity difference.  On the other
hand, the fairly strong As/P--H covalent bonds hinder H atoms from
moving to interstitial regions.  Hence H--H covalent bonds are
unlikely to form.  This may be responsible for their positive
formation enthalpies in Fig.\
\ref{electroNegativityForamtionEnthalpy}.

Once metallic hydrides are stabilized by high pressure, one can ask
whether or not they are good superconductors.  Our current results
indicate that three features of hydrides are essential for strong EPC
and a high $T_{c}$. The first is a high N(E$_{f}$).  While this
quantity certainly depends on the specific band structure and
electronic occupation, highly symmetric structures are in principle
favorable for high values of N(E$_{f}$).  Interestingly this occurs
with small magnitudes of $|\chi_{M} - \chi_{H}|$
(Fig. \ref{StatesEfermi}).  The second feature is the appearance of
substantial H-derived states in proximity to E$_{f}$, mimicking pure
metallic hydrogen.  H-rich compounds are more likely to satisfy this
criterion.  The third feature is the large electronic deformation
potential with respect to motion of H atoms.  The hydrides with strong
chemical bonding, especially covalent bonding, are more favored by
this feature.  In this regard, the $T_{c}$ of S hydrides are predicted
to be enhanced by alloying with more electronegative elements
(\textit{e.g.}, P and O) \cite{PhysRevB.92.060508, arXiv.1507.08525G}
to further strengthen the covalent bonding.  For the hydrides
containing quasi-molecular assemblies of H atoms \cite{PNAS.106.17640,
  PhysRevLett.106.237002, arXiv.1412.1542S, PNAS.109.6463,
  PhysChemChemPhys.17.19379, PNAS.107.15708, PhysRevLett.101.107002,
  PNAS.107.1317, arXiv.1503.08587L, arXiv.1503.00396Z}, the
deformation potential associated with intra-molecular H--H stretching
motion is fairly large (due to the strong H--H covalent bonds).
Unfortunately, as demonstrated in Fig.\ \ref{epc}, the corresponding
phonon modes make a quite limited contribution to the EPC.  An even
higher superconducting $T_{c}$ would be expected if these phonon modes
were involved in the EPC, for example, under further compression.

While preparing this manuscript, our attention was drawn to two
reports on superconductivity in pnictogen hydrides.  One is a
preliminary observation of high $T_{c}$ superconductivity in P
hydrides from $\sim$30 K increasing continuously to more than 100 K up
to pressures above 200 GPa \cite{arXiv.1508.06224D}.  While more
experimental data are required to verify this finding, the basic idea
coincides with the main motivation of this work.  However, based on
our structure searching results, P hydrides are predicted to be
unstable against elemental decomposition up to 400 GPa.  Our results
point to the possibility that P hydrides formed in the experiments
\cite{arXiv.1508.06224D} may be metastable and could be stabilized by
kinetic processes at high pressures.  
The conclusion that the P hydrides are thermodynamically unstable at megabar pressures
is consistent with recent theoretical reports \cite{arXiv.1509.05455S,
  arXiv.1512.02132F}.  Further analysis of the metastability, energy
barriers related to kinetic stability, as well as superconductivity
calculations for P hydrides, will be published elsewhere.  Another is
the theoretical prediction of 106 K superconductivity in SbH$_{4}$ at
150 GPa by Ma \textit{et al.}\ \cite{arXiv.1506.03889M}.  Using a
structure searching method based on a genetic algorithm
\cite{JChemPhys124.244704}, they obtained the same high-symmetry
structure (space group $P6_{3}/mmc$) of SbH$_{4}$ that we have found
in our work.  Their calculations of superconducting properties also
agree with ours.  Experimental attempts to synthesise Sb hydrides at
high pressures and explore their superconducting properties are
eagerly anticipated to test this theoretical prediction.

\section{\textbf{IV. CONCLUSIONS}}

We have used first principles structure searching methods to study the
hitherto unknown phase diagram of solid pnictogens (\textit{i.e.}, P,
As, Sb) hydrides at high pressures with the aim of predicting new
high-$T_{c}$ superconductors.  Surprisingly, except for the molecular
solid phases stabilized by weak van der Waals interactions near
ambient pressure, we did not find any P hydrides to be stable with
respect to decomposition into the elements up to 400 GPa.  The
hydrides containing the heavy element Sb are found to be the most
stable at high pressures.  This was unexpected because at ambient
pressure the lighter pnictogen hydrides are the most stable.

We have predicted several stable superconducting hydrides (AsH,
AsH$_8$, SbH, SbH$_3$ and SbH$_4$).  Among them, two H-rich compounds,
AsH$_{8}$ and SbH$_{4}$ are predicted to exhibit $T_{c}$ values of
$\sim$150 K above 350 GPa, and $\sim$100 K above 150 GPa,
respectively.  SbH$_{4}$ is energetically very stable and adopts a
highly symmetrical hexagonal structure, and the stabilization pressure
is well within reach of modern diamond anvil cell experiments.
Finally, through systematic analysis of current and previous studies
of binary hydrides, we determined a close connection between the
high-pressure behavior of hydrides and ambient pressure chemical
quantities.  In particular we found that the electronegativity
difference between the constituent elements is a good descriptor for
characterizing the structure, chemical bonding, thermodynamic
stability and superconducting properties of hydrides at high
pressures.  Analysing these trends could provide further insights into
the intriguing and diverse chemical and physical properties of
compressed hydrides.  Our work provides a useful roadmap for
discovering more stable compressed hydrides and exploration of their
superconducting properties.

\section*{\textbf{ASSOCIATED CONTENT}}
\textbf{Supporting Information.}  
Effect of zero point energy on phase stabilities,
Electronic structure, phonon, electron-phonon coupling results of H-poor compounds,
Electron localization function and Bader charge analysis results,
Explicit structural information and calculated superconducting properties of the identified stable hydrides.

\section*{\textbf{ACKNOWLEDGMENTS}}
The authors thank Eva Zurek for sharing structure data for iodine
hydride.  The work at Jilin Univ. is supported by the funding of
National Natural Science Foundation of China under Grant Nos. 11274136 and 11534003, 2012 Changjiang
Scholar of Ministry of Education and the Postdoctoral Science
Foundation of China under grant 2013M541283.  L.Z.\ acknowledges
funding support from the Recruitment Program of Global Youth Experts in China. 
Part of calculations was performed in the high performance computing center of Jilin Univ.  
R.J.N.\ acknowledges financial support
from the Engineering and Physical Sciences Research Council (EPSRC) of
the UK [EP/J017639/1].  R.J.N.\ and C.J.P.\ acknowledge use of the
Archer facility of the U.K.'s national high-performance computing
service (for which access was obtained via the UKCP consortium
[EP/K013564/1]).

\bibliography{ref_hydrides}

\begin{thebibliography}{80}
\expandafter\ifx\csname natexlab\endcsname\relax\def\natexlab#1{#1}\fi
\expandafter\ifx\csname bibnamefont\endcsname\relax
  \def\bibnamefont#1{#1}\fi
\expandafter\ifx\csname bibfnamefont\endcsname\relax
  \def\bibfnamefont#1{#1}\fi
\expandafter\ifx\csname citenamefont\endcsname\relax
  \def\citenamefont#1{#1}\fi
\expandafter\ifx\csname url\endcsname\relax
  \def\url#1{\texttt{#1}}\fi
\expandafter\ifx\csname urlprefix\endcsname\relax\def\urlprefix{URL }\fi
\providecommand{\bibinfo}[2]{#2}
\providecommand{\eprint}[2][]{\url{#2}}

\bibitem[{\citenamefont{Kamihara et~al.}(2008)\citenamefont{Kamihara, Watanabe,
  Hirano, and Hosono}}]{jacs.130.3296}
\bibinfo{author}{\bibfnamefont{Y.}~\bibnamefont{Kamihara}},
  \bibinfo{author}{\bibfnamefont{T.}~\bibnamefont{Watanabe}},
  \bibinfo{author}{\bibfnamefont{M.}~\bibnamefont{Hirano}}, \bibnamefont{and}
  \bibinfo{author}{\bibfnamefont{H.}~\bibnamefont{Hosono}},
  \bibinfo{journal}{J. Am. Chem. Soc.} \textbf{\bibinfo{volume}{130}},
  \bibinfo{pages}{3296} (\bibinfo{year}{2008}).

\bibitem[{\citenamefont{Fang et~al.}(2008)\citenamefont{Fang, Yao, Tsai, Hu,
  and Kivelson}}]{PhysRevB.77.224509}
\bibinfo{author}{\bibfnamefont{C.}~\bibnamefont{Fang}},
  \bibinfo{author}{\bibfnamefont{H.}~\bibnamefont{Yao}},
  \bibinfo{author}{\bibfnamefont{W.-F.} \bibnamefont{Tsai}},
  \bibinfo{author}{\bibfnamefont{J.}~\bibnamefont{Hu}}, \bibnamefont{and}
  \bibinfo{author}{\bibfnamefont{S.~A.} \bibnamefont{Kivelson}},
  \bibinfo{journal}{Phys. Rev. B} \textbf{\bibinfo{volume}{77}},
  \bibinfo{pages}{224509} (\bibinfo{year}{2008}).

\bibitem[{\citenamefont{Mazin et~al.}(2008)\citenamefont{Mazin, Singh,
  Johannes, and Du}}]{PhysRevLett.101.057003}
\bibinfo{author}{\bibfnamefont{I.~I.} \bibnamefont{Mazin}},
  \bibinfo{author}{\bibfnamefont{D.~J.} \bibnamefont{Singh}},
  \bibinfo{author}{\bibfnamefont{M.~D.} \bibnamefont{Johannes}},
  \bibnamefont{and} \bibinfo{author}{\bibfnamefont{M.~H.} \bibnamefont{Du}},
  \bibinfo{journal}{Phys. Rev. Lett.} \textbf{\bibinfo{volume}{101}},
  \bibinfo{pages}{057003} (\bibinfo{year}{2008}).

\bibitem[{\citenamefont{Scalapino}(2012)}]{RevModPhys.84.1383}
\bibinfo{author}{\bibfnamefont{D.~J.} \bibnamefont{Scalapino}},
  \bibinfo{journal}{Rev. Mod. Phys.} \textbf{\bibinfo{volume}{84}},
  \bibinfo{pages}{1383} (\bibinfo{year}{2012}).

\bibitem[{\citenamefont{Drozdov et~al.}(2015)\citenamefont{Drozdov, Eremets,
  Troyan, Ksenofontov, and Shylin}}]{Nature.525.73}
\bibinfo{author}{\bibfnamefont{A.~P.} \bibnamefont{Drozdov}},
  \bibinfo{author}{\bibfnamefont{M.~I.} \bibnamefont{Eremets}},
  \bibinfo{author}{\bibfnamefont{I.~A.} \bibnamefont{Troyan}},
  \bibinfo{author}{\bibfnamefont{V.}~\bibnamefont{Ksenofontov}},
  \bibnamefont{and} \bibinfo{author}{\bibfnamefont{S.~I.}
  \bibnamefont{Shylin}}, \bibinfo{journal}{Nature}
  \textbf{\bibinfo{volume}{525}}, \bibinfo{pages}{73} (\bibinfo{year}{2015}).

\bibitem[{\citenamefont{{Einaga} et~al.}(2015)\citenamefont{{Einaga}, {Sakata},
  {Ishikawa}, {Shimizu}, {Eremets}, {Drozdov}, {Troyan}, {Hirao}, and
  {Ohishi}}}]{arXiv.1509.03156E}
\bibinfo{author}{\bibfnamefont{M.}~\bibnamefont{{Einaga}}},
  \bibinfo{author}{\bibfnamefont{M.}~\bibnamefont{{Sakata}}},
  \bibinfo{author}{\bibfnamefont{T.}~\bibnamefont{{Ishikawa}}},
  \bibinfo{author}{\bibfnamefont{K.}~\bibnamefont{{Shimizu}}},
  \bibinfo{author}{\bibfnamefont{M.}~\bibnamefont{{Eremets}}},
  \bibinfo{author}{\bibfnamefont{A.}~\bibnamefont{{Drozdov}}},
  \bibinfo{author}{\bibfnamefont{I.}~\bibnamefont{{Troyan}}},
  \bibinfo{author}{\bibfnamefont{N.}~\bibnamefont{{Hirao}}}, \bibnamefont{and}
  \bibinfo{author}{\bibfnamefont{Y.}~\bibnamefont{{Ohishi}}}, p.
  \bibinfo{pages}{arXiv:1509.03156} (\bibinfo{year}{2015}).

\bibitem[{\citenamefont{Nicol and Carbotte}(2015)}]{PhysRevB.91.220507}
\bibinfo{author}{\bibfnamefont{E.~J.} \bibnamefont{Nicol}} \bibnamefont{and}
  \bibinfo{author}{\bibfnamefont{J.~P.} \bibnamefont{Carbotte}},
  \bibinfo{journal}{Phys. Rev. B} \textbf{\bibinfo{volume}{91}},
  \bibinfo{pages}{220507} (\bibinfo{year}{2015}).

\bibitem[{\citenamefont{{Bianconi} and {Jarlborg}}(2015)}]{arXiv.1507.01093B}
\bibinfo{author}{\bibfnamefont{A.}~\bibnamefont{{Bianconi}}} \bibnamefont{and}
  \bibinfo{author}{\bibfnamefont{T.}~\bibnamefont{{Jarlborg}}}, p.
  \bibinfo{pages}{arXiv:1507.01093} (\bibinfo{year}{2015}).

\bibitem[{\citenamefont{Hirsch and Marsiglio}(2015)}]{PhysicaC.511.45}
\bibinfo{author}{\bibfnamefont{J.}~\bibnamefont{Hirsch}} \bibnamefont{and}
  \bibinfo{author}{\bibfnamefont{F.}~\bibnamefont{Marsiglio}},
  \bibinfo{journal}{Phys. C} \textbf{\bibinfo{volume}{511}},
  \bibinfo{pages}{45} (\bibinfo{year}{2015}).

\bibitem[{\citenamefont{Bernstein et~al.}(2015)\citenamefont{Bernstein,
  Hellberg, Johannes, Mazin, and Mehl}}]{PhysRevB.91.060511}
\bibinfo{author}{\bibfnamefont{N.}~\bibnamefont{Bernstein}},
  \bibinfo{author}{\bibfnamefont{C.~S.} \bibnamefont{Hellberg}},
  \bibinfo{author}{\bibfnamefont{M.~D.} \bibnamefont{Johannes}},
  \bibinfo{author}{\bibfnamefont{I.~I.} \bibnamefont{Mazin}}, \bibnamefont{and}
  \bibinfo{author}{\bibfnamefont{M.~J.} \bibnamefont{Mehl}},
  \bibinfo{journal}{Phys. Rev. B} \textbf{\bibinfo{volume}{91}},
  \bibinfo{pages}{060511} (\bibinfo{year}{2015}).

\bibitem[{\citenamefont{Heil and Boeri}(2015)}]{PhysRevB.92.060508}
\bibinfo{author}{\bibfnamefont{C.}~\bibnamefont{Heil}} \bibnamefont{and}
  \bibinfo{author}{\bibfnamefont{L.}~\bibnamefont{Boeri}},
  \bibinfo{journal}{Phys. Rev. B} \textbf{\bibinfo{volume}{92}},
  \bibinfo{pages}{060508} (\bibinfo{year}{2015}).

\bibitem[{\citenamefont{Errea et~al.}(2015)\citenamefont{Errea, Calandra,
  Pickard, Nelson, Needs, Li, Liu, Zhang, Ma, and
  Mauri}}]{PhysRevLett.114.157004}
\bibinfo{author}{\bibfnamefont{I.}~\bibnamefont{Errea}},
  \bibinfo{author}{\bibfnamefont{M.}~\bibnamefont{Calandra}},
  \bibinfo{author}{\bibfnamefont{C.~J.} \bibnamefont{Pickard}},
  \bibinfo{author}{\bibfnamefont{J.}~\bibnamefont{Nelson}},
  \bibinfo{author}{\bibfnamefont{R.~J.} \bibnamefont{Needs}},
  \bibinfo{author}{\bibfnamefont{Y.}~\bibnamefont{Li}},
  \bibinfo{author}{\bibfnamefont{H.}~\bibnamefont{Liu}},
  \bibinfo{author}{\bibfnamefont{Y.}~\bibnamefont{Zhang}},
  \bibinfo{author}{\bibfnamefont{Y.}~\bibnamefont{Ma}}, \bibnamefont{and}
  \bibinfo{author}{\bibfnamefont{F.}~\bibnamefont{Mauri}},
  \bibinfo{journal}{Phys. Rev. Lett.} \textbf{\bibinfo{volume}{114}},
  \bibinfo{pages}{157004} (\bibinfo{year}{2015}).

\bibitem[{\citenamefont{Ashcroft}(1968)}]{PhysRevLett.21.1748}
\bibinfo{author}{\bibfnamefont{N.~W.} \bibnamefont{Ashcroft}},
  \bibinfo{journal}{Phys. Rev. Lett.} \textbf{\bibinfo{volume}{21}},
  \bibinfo{pages}{1748} (\bibinfo{year}{1968}).

\bibitem[{\citenamefont{Ashcroft}(2004)}]{PhysRevLett.92.187002}
\bibinfo{author}{\bibfnamefont{N.~W.} \bibnamefont{Ashcroft}},
  \bibinfo{journal}{Phys. Rev. Lett.} \textbf{\bibinfo{volume}{92}},
  \bibinfo{pages}{187002} (\bibinfo{year}{2004}).

\bibitem[{\citenamefont{Feng et~al.}(2006)\citenamefont{Feng, Grochala,
  Jaro\ifmmode~\acute{n}\else \'{n}\fi{}, Hoffmann, Bergara, and
  Ashcroft}}]{PhysRevLett.96.017006}
\bibinfo{author}{\bibfnamefont{J.}~\bibnamefont{Feng}},
  \bibinfo{author}{\bibfnamefont{W.}~\bibnamefont{Grochala}},
  \bibinfo{author}{\bibfnamefont{T.}~\bibnamefont{Jaro\ifmmode~\acute{n}\else
  \'{n}\fi{}}}, \bibinfo{author}{\bibfnamefont{R.}~\bibnamefont{Hoffmann}},
  \bibinfo{author}{\bibfnamefont{A.}~\bibnamefont{Bergara}}, \bibnamefont{and}
  \bibinfo{author}{\bibfnamefont{N.~W.} \bibnamefont{Ashcroft}},
  \bibinfo{journal}{Phys. Rev. Lett.} \textbf{\bibinfo{volume}{96}},
  \bibinfo{pages}{017006} (\bibinfo{year}{2006}).

\bibitem[{\citenamefont{Li et~al.}(2014)\citenamefont{Li, Hao, Liu, Li, and
  Ma}}]{JCP.140.174712}
\bibinfo{author}{\bibfnamefont{Y.}~\bibnamefont{Li}},
  \bibinfo{author}{\bibfnamefont{J.}~\bibnamefont{Hao}},
  \bibinfo{author}{\bibfnamefont{H.}~\bibnamefont{Liu}},
  \bibinfo{author}{\bibfnamefont{Y.}~\bibnamefont{Li}}, \bibnamefont{and}
  \bibinfo{author}{\bibfnamefont{Y.}~\bibnamefont{Ma}}, \bibinfo{journal}{J.
  Chem. Phys.} \textbf{\bibinfo{volume}{140}}, \bibinfo{pages}{174712}
  (\bibinfo{year}{2014}).

\bibitem[{\citenamefont{Duan et~al.}(2014)\citenamefont{Duan, Liu, Tian, Li,
  Huang, Zhao, Yu, Liu, Tian, and Cui}}]{ScientificReports.4}
\bibinfo{author}{\bibfnamefont{D.}~\bibnamefont{Duan}},
  \bibinfo{author}{\bibfnamefont{Y.}~\bibnamefont{Liu}},
  \bibinfo{author}{\bibfnamefont{F.}~\bibnamefont{Tian}},
  \bibinfo{author}{\bibfnamefont{D.}~\bibnamefont{Li}},
  \bibinfo{author}{\bibfnamefont{X.}~\bibnamefont{Huang}},
  \bibinfo{author}{\bibfnamefont{Z.}~\bibnamefont{Zhao}},
  \bibinfo{author}{\bibfnamefont{H.}~\bibnamefont{Yu}},
  \bibinfo{author}{\bibfnamefont{B.}~\bibnamefont{Liu}},
  \bibinfo{author}{\bibfnamefont{W.}~\bibnamefont{Tian}}, \bibnamefont{and}
  \bibinfo{author}{\bibfnamefont{T.}~\bibnamefont{Cui}}, \bibinfo{journal}{Sci.
  Rep.} \textbf{\bibinfo{volume}{4}}, \bibinfo{pages}{6968}
  (\bibinfo{year}{2014}).

\bibitem[{\citenamefont{Papaconstantopoulos
  et~al.}(2015)\citenamefont{Papaconstantopoulos, Klein, Mehl, and
  Pickett}}]{PhysRevB.91.184511}
\bibinfo{author}{\bibfnamefont{D.~A.} \bibnamefont{Papaconstantopoulos}},
  \bibinfo{author}{\bibfnamefont{B.~M.} \bibnamefont{Klein}},
  \bibinfo{author}{\bibfnamefont{M.~J.} \bibnamefont{Mehl}}, \bibnamefont{and}
  \bibinfo{author}{\bibfnamefont{W.~E.} \bibnamefont{Pickett}},
  \bibinfo{journal}{Phys. Rev. B} \textbf{\bibinfo{volume}{91}},
  \bibinfo{pages}{184511} (\bibinfo{year}{2015}).

\bibitem[{\citenamefont{{Flores-Livas}
  et~al.}(2015{\natexlab{a}})\citenamefont{{Flores-Livas}, {Sanna}, and
  {Gross}}}]{arXiv.1501.06336F}
\bibinfo{author}{\bibfnamefont{J.~A.} \bibnamefont{{Flores-Livas}}},
  \bibinfo{author}{\bibfnamefont{A.}~\bibnamefont{{Sanna}}}, \bibnamefont{and}
  \bibinfo{author}{\bibfnamefont{E.~K.~U.} \bibnamefont{{Gross}}}, p.
  \bibinfo{pages}{arXiv:1501.06336} (\bibinfo{year}{2015}{\natexlab{a}}).

\bibitem[{\citenamefont{Wang and Ma}(2014)}]{JCP.140.040901}
\bibinfo{author}{\bibfnamefont{Y.}~\bibnamefont{Wang}} \bibnamefont{and}
  \bibinfo{author}{\bibfnamefont{Y.}~\bibnamefont{Ma}}, \bibinfo{journal}{J.
  Chem. Phys.} \textbf{\bibinfo{volume}{140}}, \bibinfo{pages}{040901}
  (\bibinfo{year}{2014}).

\bibitem[{\citenamefont{Duan et~al.}(2015)\citenamefont{Duan, Huang, Tian, Li,
  Yu, Liu, Ma, Liu, and Cui}}]{PhysRevB.91.180502}
\bibinfo{author}{\bibfnamefont{D.}~\bibnamefont{Duan}},
  \bibinfo{author}{\bibfnamefont{X.}~\bibnamefont{Huang}},
  \bibinfo{author}{\bibfnamefont{F.}~\bibnamefont{Tian}},
  \bibinfo{author}{\bibfnamefont{D.}~\bibnamefont{Li}},
  \bibinfo{author}{\bibfnamefont{H.}~\bibnamefont{Yu}},
  \bibinfo{author}{\bibfnamefont{Y.}~\bibnamefont{Liu}},
  \bibinfo{author}{\bibfnamefont{Y.}~\bibnamefont{Ma}},
  \bibinfo{author}{\bibfnamefont{B.}~\bibnamefont{Liu}}, \bibnamefont{and}
  \bibinfo{author}{\bibfnamefont{T.}~\bibnamefont{Cui}},
  \bibinfo{journal}{Phys. Rev. B} \textbf{\bibinfo{volume}{91}},
  \bibinfo{pages}{180502} (\bibinfo{year}{2015}).

\bibitem[{\citenamefont{{Zhang} et~al.}(2015)\citenamefont{{Zhang}, {Wang},
  {Zhang}, {Liu}, {Zhong}, {Song}, {Yang}, {Zhang}, and
  {Ma}}}]{arXiv.1502.02607Z}
\bibinfo{author}{\bibfnamefont{S.}~\bibnamefont{{Zhang}}},
  \bibinfo{author}{\bibfnamefont{Y.}~\bibnamefont{{Wang}}},
  \bibinfo{author}{\bibfnamefont{J.}~\bibnamefont{{Zhang}}},
  \bibinfo{author}{\bibfnamefont{H.}~\bibnamefont{{Liu}}},
  \bibinfo{author}{\bibfnamefont{X.}~\bibnamefont{{Zhong}}},
  \bibinfo{author}{\bibfnamefont{H.-F.} \bibnamefont{{Song}}},
  \bibinfo{author}{\bibfnamefont{G.}~\bibnamefont{{Yang}}},
  \bibinfo{author}{\bibfnamefont{L.}~\bibnamefont{{Zhang}}}, \bibnamefont{and}
  \bibinfo{author}{\bibfnamefont{Y.}~\bibnamefont{{Ma}}}, p.
  \bibinfo{pages}{arXiv:1502.02607} (\bibinfo{year}{2015}).

\bibitem[{\citenamefont{{Zhong} et~al.}(2015)\citenamefont{{Zhong}, {Wang},
  {Zhang}, {Liu}, {Zhang}, {Song}, {Yang}, {Zhang}, and
  {Ma}}}]{arXiv.1503.00396Z}
\bibinfo{author}{\bibfnamefont{X.}~\bibnamefont{{Zhong}}},
  \bibinfo{author}{\bibfnamefont{H.}~\bibnamefont{{Wang}}},
  \bibinfo{author}{\bibfnamefont{J.}~\bibnamefont{{Zhang}}},
  \bibinfo{author}{\bibfnamefont{H.}~\bibnamefont{{Liu}}},
  \bibinfo{author}{\bibfnamefont{S.}~\bibnamefont{{Zhang}}},
  \bibinfo{author}{\bibfnamefont{H.-F.} \bibnamefont{{Song}}},
  \bibinfo{author}{\bibfnamefont{G.}~\bibnamefont{{Yang}}},
  \bibinfo{author}{\bibfnamefont{L.}~\bibnamefont{{Zhang}}}, \bibnamefont{and}
  \bibinfo{author}{\bibfnamefont{Y.}~\bibnamefont{{Ma}}}, p.
  \bibinfo{pages}{arXiv:1503.00396} (\bibinfo{year}{2015}).

\bibitem[{\citenamefont{Natta and
  Casazza}(1930)}]{GazzettaChimicaItaliana.60.981}
\bibinfo{author}{\bibfnamefont{G.}~\bibnamefont{Natta}} \bibnamefont{and}
  \bibinfo{author}{\bibfnamefont{E.}~\bibnamefont{Casazza}},
  \bibinfo{journal}{Gazz. Chim. Ital.} \textbf{\bibinfo{volume}{60}},
  \bibinfo{pages}{981} (\bibinfo{year}{1930}).

\bibitem[{\citenamefont{Sennikov}(1994)}]{JPC.98.4973}
\bibinfo{author}{\bibfnamefont{P.}~\bibnamefont{Sennikov}},
  \bibinfo{journal}{J. Phys. Chem.} \textbf{\bibinfo{volume}{98}},
  \bibinfo{pages}{4973} (\bibinfo{year}{1994}).

\bibitem[{\citenamefont{Baudler and Glinka}(1993)}]{ChemicalReviews.93.1623}
\bibinfo{author}{\bibfnamefont{M.}~\bibnamefont{Baudler}} \bibnamefont{and}
  \bibinfo{author}{\bibfnamefont{K.}~\bibnamefont{Glinka}},
  \bibinfo{journal}{Chem. Rev.} \textbf{\bibinfo{volume}{93}},
  \bibinfo{pages}{1623} (\bibinfo{year}{1993}).

\bibitem[{\citenamefont{Holleman and Wiberg}(2001)}]{InorganicChemistry}
\bibinfo{author}{\bibfnamefont{A.}~\bibnamefont{Holleman}} \bibnamefont{and}
  \bibinfo{author}{\bibfnamefont{E.}~\bibnamefont{Wiberg}},
  \emph{\bibinfo{title}{Inorganic Chemistry}} (\bibinfo{publisher}{Academic
  Press}, \bibinfo{address}{M$\ddot u$nchen}, \bibinfo{year}{2001}).

\bibitem[{\citenamefont{Matthias}(1957)}]{matthias1957progress}
\bibinfo{author}{\bibfnamefont{B.}~\bibnamefont{Matthias}},
  \emph{\bibinfo{title}{Progress in low temperature physics}},
  vol.~\bibinfo{volume}{2} (\bibinfo{publisher}{North Holland Publishing
  Company}, \bibinfo{address}{Leiden}, \bibinfo{year}{1957}).

\bibitem[{\citenamefont{Pines}(1958)}]{PhysRev.109.280}
\bibinfo{author}{\bibfnamefont{D.}~\bibnamefont{Pines}},
  \bibinfo{journal}{Phys. Rev.} \textbf{\bibinfo{volume}{109}},
  \bibinfo{pages}{280} (\bibinfo{year}{1958}).

\bibitem[{\citenamefont{Wang et~al.}(2012{\natexlab{a}})\citenamefont{Wang,
  Tse, Tanaka, Iitaka, and Ma}}]{PNAS.109.6463}
\bibinfo{author}{\bibfnamefont{H.}~\bibnamefont{Wang}},
  \bibinfo{author}{\bibfnamefont{J.~S.} \bibnamefont{Tse}},
  \bibinfo{author}{\bibfnamefont{K.}~\bibnamefont{Tanaka}},
  \bibinfo{author}{\bibfnamefont{T.}~\bibnamefont{Iitaka}}, \bibnamefont{and}
  \bibinfo{author}{\bibfnamefont{Y.}~\bibnamefont{Ma}}, \bibinfo{journal}{Proc.
  Natl. Acad. Sci. U.S.A.} \textbf{\bibinfo{volume}{109}},
  \bibinfo{pages}{6463} (\bibinfo{year}{2012}{\natexlab{a}}).

\bibitem[{\citenamefont{Li et~al.}(2010)\citenamefont{Li, Gao, Xie, Ma, Cui,
  and Zou}}]{PNAS.107.15708}
\bibinfo{author}{\bibfnamefont{Y.}~\bibnamefont{Li}},
  \bibinfo{author}{\bibfnamefont{G.}~\bibnamefont{Gao}},
  \bibinfo{author}{\bibfnamefont{Y.}~\bibnamefont{Xie}},
  \bibinfo{author}{\bibfnamefont{Y.}~\bibnamefont{Ma}},
  \bibinfo{author}{\bibfnamefont{T.}~\bibnamefont{Cui}}, \bibnamefont{and}
  \bibinfo{author}{\bibfnamefont{G.}~\bibnamefont{Zou}},
  \bibinfo{journal}{Proc. Natl. Acad. Sci. U.S.A.}
  \textbf{\bibinfo{volume}{107}}, \bibinfo{pages}{15708}
  (\bibinfo{year}{2010}).

\bibitem[{\citenamefont{Gao et~al.}(2008)\citenamefont{Gao, Oganov, Bergara,
  Martinez-Canales, Cui, Iitaka, Ma, and Zou}}]{PhysRevLett.101.107002}
\bibinfo{author}{\bibfnamefont{G.}~\bibnamefont{Gao}},
  \bibinfo{author}{\bibfnamefont{A.~R.} \bibnamefont{Oganov}},
  \bibinfo{author}{\bibfnamefont{A.}~\bibnamefont{Bergara}},
  \bibinfo{author}{\bibfnamefont{M.}~\bibnamefont{Martinez-Canales}},
  \bibinfo{author}{\bibfnamefont{T.}~\bibnamefont{Cui}},
  \bibinfo{author}{\bibfnamefont{T.}~\bibnamefont{Iitaka}},
  \bibinfo{author}{\bibfnamefont{Y.}~\bibnamefont{Ma}}, \bibnamefont{and}
  \bibinfo{author}{\bibfnamefont{G.}~\bibnamefont{Zou}},
  \bibinfo{journal}{Phys. Rev. Lett.} \textbf{\bibinfo{volume}{101}},
  \bibinfo{pages}{107002} (\bibinfo{year}{2008}).

\bibitem[{\citenamefont{Gao et~al.}(2010)\citenamefont{Gao, Oganov, Li, Li,
  Wang, Cui, Ma, Bergara, Lyakhov, Iitaka et~al.}}]{PNAS.107.1317}
\bibinfo{author}{\bibfnamefont{G.}~\bibnamefont{Gao}},
  \bibinfo{author}{\bibfnamefont{A.~R.} \bibnamefont{Oganov}},
  \bibinfo{author}{\bibfnamefont{P.}~\bibnamefont{Li}},
  \bibinfo{author}{\bibfnamefont{Z.}~\bibnamefont{Li}},
  \bibinfo{author}{\bibfnamefont{H.}~\bibnamefont{Wang}},
  \bibinfo{author}{\bibfnamefont{T.}~\bibnamefont{Cui}},
  \bibinfo{author}{\bibfnamefont{Y.}~\bibnamefont{Ma}},
  \bibinfo{author}{\bibfnamefont{A.}~\bibnamefont{Bergara}},
  \bibinfo{author}{\bibfnamefont{A.~O.} \bibnamefont{Lyakhov}},
  \bibinfo{author}{\bibfnamefont{T.}~\bibnamefont{Iitaka}},
  \bibnamefont{et~al.}, \bibinfo{journal}{Proc. Natl. Acad. Sci. U.S.A.}
  \textbf{\bibinfo{volume}{107}}, \bibinfo{pages}{1317} (\bibinfo{year}{2010}).

\bibitem[{\citenamefont{{Liu} et~al.}(2015)\citenamefont{{Liu}, {Duan}, {Tian},
  {Li}, {Sha}, {Zhao}, {Zhang}, {Wu}, {Yu}, {Liu} et~al.}}]{arXiv.1503.08587L}
\bibinfo{author}{\bibfnamefont{Y.}~\bibnamefont{{Liu}}},
  \bibinfo{author}{\bibfnamefont{D.}~\bibnamefont{{Duan}}},
  \bibinfo{author}{\bibfnamefont{F.}~\bibnamefont{{Tian}}},
  \bibinfo{author}{\bibfnamefont{D.}~\bibnamefont{{Li}}},
  \bibinfo{author}{\bibfnamefont{X.}~\bibnamefont{{Sha}}},
  \bibinfo{author}{\bibfnamefont{Z.}~\bibnamefont{{Zhao}}},
  \bibinfo{author}{\bibfnamefont{H.}~\bibnamefont{{Zhang}}},
  \bibinfo{author}{\bibfnamefont{G.}~\bibnamefont{{Wu}}},
  \bibinfo{author}{\bibfnamefont{H.}~\bibnamefont{{Yu}}},
  \bibinfo{author}{\bibfnamefont{B.}~\bibnamefont{{Liu}}},
  \bibnamefont{et~al.}, p. \bibinfo{pages}{arXiv:1503.08587}
  (\bibinfo{year}{2015}).

\bibitem[{\citenamefont{{Shamp} and {Zurek}}(2015)}]{arXiv.1507.02616S}
\bibinfo{author}{\bibfnamefont{A.}~\bibnamefont{{Shamp}}} \bibnamefont{and}
  \bibinfo{author}{\bibfnamefont{E.}~\bibnamefont{{Zurek}}}, p.
  \bibinfo{pages}{arXiv:1507.02616} (\bibinfo{year}{2015}).

\bibitem[{\citenamefont{{Duan} et~al.}(2015)\citenamefont{{Duan}, {Tian},
  {Huang}, {Li}, {Yu}, {Liu}, {Ma}, {Liu}, and {Cui}}}]{arXiv.1504.01196D}
\bibinfo{author}{\bibfnamefont{D.}~\bibnamefont{{Duan}}},
  \bibinfo{author}{\bibfnamefont{F.}~\bibnamefont{{Tian}}},
  \bibinfo{author}{\bibfnamefont{X.}~\bibnamefont{{Huang}}},
  \bibinfo{author}{\bibfnamefont{D.}~\bibnamefont{{Li}}},
  \bibinfo{author}{\bibfnamefont{H.}~\bibnamefont{{Yu}}},
  \bibinfo{author}{\bibfnamefont{Y.}~\bibnamefont{{Liu}}},
  \bibinfo{author}{\bibfnamefont{Y.}~\bibnamefont{{Ma}}},
  \bibinfo{author}{\bibfnamefont{B.}~\bibnamefont{{Liu}}}, \bibnamefont{and}
  \bibinfo{author}{\bibfnamefont{T.}~\bibnamefont{{Cui}}}, p.
  \bibinfo{pages}{arXiv:1504.01196} (\bibinfo{year}{2015}).

\bibitem[{\citenamefont{Wang et~al.}(2014)\citenamefont{Wang, Yao, Zhu, Liu,
  Iitaka, Wang, and Ma}}]{JChemPhys.140.124707}
\bibinfo{author}{\bibfnamefont{Z.}~\bibnamefont{Wang}},
  \bibinfo{author}{\bibfnamefont{Y.}~\bibnamefont{Yao}},
  \bibinfo{author}{\bibfnamefont{L.}~\bibnamefont{Zhu}},
  \bibinfo{author}{\bibfnamefont{H.}~\bibnamefont{Liu}},
  \bibinfo{author}{\bibfnamefont{T.}~\bibnamefont{Iitaka}},
  \bibinfo{author}{\bibfnamefont{H.}~\bibnamefont{Wang}}, \bibnamefont{and}
  \bibinfo{author}{\bibfnamefont{Y.}~\bibnamefont{Ma}}, \bibinfo{journal}{J.
  Chem. Phys.} \textbf{\bibinfo{volume}{140}}, \bibinfo{pages}{124707}
  (\bibinfo{year}{2014}).

\bibitem[{\citenamefont{Hu et~al.}(2013)\citenamefont{Hu, Oganov, Zhu, Qian,
  Frapper, Lyakhov, and Zhou}}]{PhysRevLett.110.165504}
\bibinfo{author}{\bibfnamefont{C.-H.} \bibnamefont{Hu}},
  \bibinfo{author}{\bibfnamefont{A.~R.} \bibnamefont{Oganov}},
  \bibinfo{author}{\bibfnamefont{Q.}~\bibnamefont{Zhu}},
  \bibinfo{author}{\bibfnamefont{G.-R.} \bibnamefont{Qian}},
  \bibinfo{author}{\bibfnamefont{G.}~\bibnamefont{Frapper}},
  \bibinfo{author}{\bibfnamefont{A.~O.} \bibnamefont{Lyakhov}},
  \bibnamefont{and} \bibinfo{author}{\bibfnamefont{H.-Y.} \bibnamefont{Zhou}},
  \bibinfo{journal}{Phys. Rev. Lett.} \textbf{\bibinfo{volume}{110}},
  \bibinfo{pages}{165504} (\bibinfo{year}{2013}).

\bibitem[{\citenamefont{Lu et~al.}(2015)\citenamefont{Lu, Wu, Liu, Tse, and
  Yang}}]{RSCAdv.5.45812}
\bibinfo{author}{\bibfnamefont{S.}~\bibnamefont{Lu}},
  \bibinfo{author}{\bibfnamefont{M.}~\bibnamefont{Wu}},
  \bibinfo{author}{\bibfnamefont{H.}~\bibnamefont{Liu}},
  \bibinfo{author}{\bibfnamefont{J.~S.} \bibnamefont{Tse}}, \bibnamefont{and}
  \bibinfo{author}{\bibfnamefont{B.}~\bibnamefont{Yang}}, \bibinfo{journal}{RSC
  Adv.} \textbf{\bibinfo{volume}{5}}, \bibinfo{pages}{45812}
  (\bibinfo{year}{2015}).

\bibitem[{\citenamefont{Gao et~al.}(1994)\citenamefont{Gao, Xue, Chen, Xiong,
  Meng, Ramirez, Chu, Eggert, and Mao}}]{PhysRevB.50.4260}
\bibinfo{author}{\bibfnamefont{L.}~\bibnamefont{Gao}},
  \bibinfo{author}{\bibfnamefont{Y.~Y.} \bibnamefont{Xue}},
  \bibinfo{author}{\bibfnamefont{F.}~\bibnamefont{Chen}},
  \bibinfo{author}{\bibfnamefont{Q.}~\bibnamefont{Xiong}},
  \bibinfo{author}{\bibfnamefont{R.~L.} \bibnamefont{Meng}},
  \bibinfo{author}{\bibfnamefont{D.}~\bibnamefont{Ramirez}},
  \bibinfo{author}{\bibfnamefont{C.~W.} \bibnamefont{Chu}},
  \bibinfo{author}{\bibfnamefont{J.~H.} \bibnamefont{Eggert}},
  \bibnamefont{and} \bibinfo{author}{\bibfnamefont{H.~K.} \bibnamefont{Mao}},
  \bibinfo{journal}{Phys. Rev. B} \textbf{\bibinfo{volume}{50}},
  \bibinfo{pages}{4260} (\bibinfo{year}{1994}).

\bibitem[{\citenamefont{Eremets et~al.}(2008)\citenamefont{Eremets, Trojan,
  Medvedev, Tse, and Yao}}]{Science.319.1506}
\bibinfo{author}{\bibfnamefont{M.~I.} \bibnamefont{Eremets}},
  \bibinfo{author}{\bibfnamefont{I.~A.} \bibnamefont{Trojan}},
  \bibinfo{author}{\bibfnamefont{S.~A.} \bibnamefont{Medvedev}},
  \bibinfo{author}{\bibfnamefont{J.~S.} \bibnamefont{Tse}}, \bibnamefont{and}
  \bibinfo{author}{\bibfnamefont{Y.}~\bibnamefont{Yao}},
  \bibinfo{journal}{Science} \textbf{\bibinfo{volume}{319}},
  \bibinfo{pages}{1506} (\bibinfo{year}{2008}).

\bibitem[{\citenamefont{Degtyareva et~al.}(2009)\citenamefont{Degtyareva,
  Proctor, Guillaume, Gregoryanz, and Hanfland}}]{Degtyareva2009}
\bibinfo{author}{\bibfnamefont{O.}~\bibnamefont{Degtyareva}},
  \bibinfo{author}{\bibfnamefont{J.~E.} \bibnamefont{Proctor}},
  \bibinfo{author}{\bibfnamefont{C.~L.} \bibnamefont{Guillaume}},
  \bibinfo{author}{\bibfnamefont{E.}~\bibnamefont{Gregoryanz}},
  \bibnamefont{and} \bibinfo{author}{\bibfnamefont{M.}~\bibnamefont{Hanfland}},
  \bibinfo{journal}{Solid State Commun.} \textbf{\bibinfo{volume}{149}},
  \bibinfo{pages}{1583} (\bibinfo{year}{2009}).

\bibitem[{\citenamefont{Szczesniak and Zemia}(2015)}]{Dszc2015}
\bibinfo{author}{\bibfnamefont{D.}~\bibnamefont{Szczesniak}} \bibnamefont{and}
  \bibinfo{author}{\bibfnamefont{T.~P.} \bibnamefont{Zemia}},
  \bibinfo{journal}{Supercond. Sci. Technol.} \textbf{\bibinfo{volume}{28}},
  \bibinfo{pages}{085018} (\bibinfo{year}{2015}).

\bibitem[{\citenamefont{Wang et~al.}(2010)\citenamefont{Wang, Lv, Zhu, and
  Ma}}]{PhysRevB.82.094116}
\bibinfo{author}{\bibfnamefont{Y.}~\bibnamefont{Wang}},
  \bibinfo{author}{\bibfnamefont{J.}~\bibnamefont{Lv}},
  \bibinfo{author}{\bibfnamefont{L.}~\bibnamefont{Zhu}}, \bibnamefont{and}
  \bibinfo{author}{\bibfnamefont{Y.}~\bibnamefont{Ma}}, \bibinfo{journal}{Phys.
  Rev. B} \textbf{\bibinfo{volume}{82}}, \bibinfo{pages}{094116}
  (\bibinfo{year}{2010}).

\bibitem[{\citenamefont{Wang et~al.}(2012{\natexlab{b}})\citenamefont{Wang, Lv,
  Zhu, and Ma}}]{CPC.183.2063}
\bibinfo{author}{\bibfnamefont{Y.}~\bibnamefont{Wang}},
  \bibinfo{author}{\bibfnamefont{J.}~\bibnamefont{Lv}},
  \bibinfo{author}{\bibfnamefont{L.}~\bibnamefont{Zhu}}, \bibnamefont{and}
  \bibinfo{author}{\bibfnamefont{Y.}~\bibnamefont{Ma}},
  \bibinfo{journal}{Comput. Phys. Commun.} \textbf{\bibinfo{volume}{183}},
  \bibinfo{pages}{2063} (\bibinfo{year}{2012}{\natexlab{b}}).

\bibitem[{\citenamefont{Pickard and Needs}(2006)}]{PhysRevLett.97.045504}
\bibinfo{author}{\bibfnamefont{C.~J.} \bibnamefont{Pickard}} \bibnamefont{and}
  \bibinfo{author}{\bibfnamefont{R.~J.} \bibnamefont{Needs}},
  \bibinfo{journal}{Phys. Rev. Lett.} \textbf{\bibinfo{volume}{97}},
  \bibinfo{pages}{045504} (\bibinfo{year}{2006}).

\bibitem[{\citenamefont{Pickard and
  Needs}(2011)}]{JPhysCondensMatter.23.053201}
\bibinfo{author}{\bibfnamefont{C.~J.} \bibnamefont{Pickard}} \bibnamefont{and}
  \bibinfo{author}{\bibfnamefont{R.~J.} \bibnamefont{Needs}},
  \bibinfo{journal}{J. Phys. Condens. Matter} \textbf{\bibinfo{volume}{23}},
  \bibinfo{pages}{053201} (\bibinfo{year}{2011}).

\bibitem[{\citenamefont{Kresse and Furthm\"uller}(1996)}]{PhysRevB.54.11169}
\bibinfo{author}{\bibfnamefont{G.}~\bibnamefont{Kresse}} \bibnamefont{and}
  \bibinfo{author}{\bibfnamefont{J.}~\bibnamefont{Furthm\"uller}},
  \bibinfo{journal}{Phys. Rev. B} \textbf{\bibinfo{volume}{54}},
  \bibinfo{pages}{11169} (\bibinfo{year}{1996}).

\bibitem[{\citenamefont{Bl\"ochl}(1994)}]{PhysRevB.50.17953}
\bibinfo{author}{\bibfnamefont{P.~E.} \bibnamefont{Bl\"ochl}},
  \bibinfo{journal}{Phys. Rev. B} \textbf{\bibinfo{volume}{50}},
  \bibinfo{pages}{17953} (\bibinfo{year}{1994}).

\bibitem[{\citenamefont{Perdew et~al.}(1992)\citenamefont{Perdew, Chevary,
  Vosko, Jackson, Pederson, Singh, and Fiolhais}}]{PhysRevB.46.6671}
\bibinfo{author}{\bibfnamefont{J.~P.} \bibnamefont{Perdew}},
  \bibinfo{author}{\bibfnamefont{J.~A.} \bibnamefont{Chevary}},
  \bibinfo{author}{\bibfnamefont{S.~H.} \bibnamefont{Vosko}},
  \bibinfo{author}{\bibfnamefont{K.~A.} \bibnamefont{Jackson}},
  \bibinfo{author}{\bibfnamefont{M.~R.} \bibnamefont{Pederson}},
  \bibinfo{author}{\bibfnamefont{D.~J.} \bibnamefont{Singh}}, \bibnamefont{and}
  \bibinfo{author}{\bibfnamefont{C.}~\bibnamefont{Fiolhais}},
  \bibinfo{journal}{Phys. Rev. B} \textbf{\bibinfo{volume}{46}},
  \bibinfo{pages}{6671} (\bibinfo{year}{1992}).

\bibitem[{\citenamefont{Giannozzi et~al.}(2009)\citenamefont{Giannozzi, Baroni,
  Bonini, Calandra, Car, Cavazzoni, Ceresoli, Chiarotti, Cococcioni, Dabo
  et~al.}}]{JPhysCondensMatter.21.395502}
\bibinfo{author}{\bibfnamefont{P.}~\bibnamefont{Giannozzi}},
  \bibinfo{author}{\bibfnamefont{S.}~\bibnamefont{Baroni}},
  \bibinfo{author}{\bibfnamefont{N.}~\bibnamefont{Bonini}},
  \bibinfo{author}{\bibfnamefont{M.}~\bibnamefont{Calandra}},
  \bibinfo{author}{\bibfnamefont{R.}~\bibnamefont{Car}},
  \bibinfo{author}{\bibfnamefont{C.}~\bibnamefont{Cavazzoni}},
  \bibinfo{author}{\bibfnamefont{D.}~\bibnamefont{Ceresoli}},
  \bibinfo{author}{\bibfnamefont{G.~L.} \bibnamefont{Chiarotti}},
  \bibinfo{author}{\bibfnamefont{M.}~\bibnamefont{Cococcioni}},
  \bibinfo{author}{\bibfnamefont{I.}~\bibnamefont{Dabo}}, \bibnamefont{et~al.},
  \bibinfo{journal}{J. Phys. Condens. Matter} \textbf{\bibinfo{volume}{21}},
  \bibinfo{pages}{395502} (\bibinfo{year}{2009}).

\bibitem[{\citenamefont{Methfessel and Paxton}(1989)}]{PhysRevB.40.3616}
\bibinfo{author}{\bibfnamefont{M.}~\bibnamefont{Methfessel}} \bibnamefont{and}
  \bibinfo{author}{\bibfnamefont{A.~T.} \bibnamefont{Paxton}},
  \bibinfo{journal}{Phys. Rev. B} \textbf{\bibinfo{volume}{40}},
  \bibinfo{pages}{3616} (\bibinfo{year}{1989}).

\bibitem[{\citenamefont{Katzke and Tol\'edano}(2008)}]{PhysRevB.77.024109}
\bibinfo{author}{\bibfnamefont{H.}~\bibnamefont{Katzke}} \bibnamefont{and}
  \bibinfo{author}{\bibfnamefont{P.}~\bibnamefont{Tol\'edano}},
  \bibinfo{journal}{Phys. Rev. B} \textbf{\bibinfo{volume}{77}},
  \bibinfo{pages}{024109} (\bibinfo{year}{2008}).

\bibitem[{\citenamefont{Pickard and Needs}(2007)}]{NaturePhysics.3.473}
\bibinfo{author}{\bibfnamefont{C.~J.} \bibnamefont{Pickard}} \bibnamefont{and}
  \bibinfo{author}{\bibfnamefont{R.~J.} \bibnamefont{Needs}},
  \bibinfo{journal}{Nat. Phys.} \textbf{\bibinfo{volume}{3}},
  \bibinfo{pages}{473} (\bibinfo{year}{2007}).

\bibitem[{\citenamefont{Zurek et~al.}(2009)\citenamefont{Zurek, Hoffmann,
  Ashcroft, Oganov, and Lyakhov}}]{PNAS.106.17640}
\bibinfo{author}{\bibfnamefont{E.}~\bibnamefont{Zurek}},
  \bibinfo{author}{\bibfnamefont{R.}~\bibnamefont{Hoffmann}},
  \bibinfo{author}{\bibfnamefont{N.~W.} \bibnamefont{Ashcroft}},
  \bibinfo{author}{\bibfnamefont{A.~R.} \bibnamefont{Oganov}},
  \bibnamefont{and} \bibinfo{author}{\bibfnamefont{A.~O.}
  \bibnamefont{Lyakhov}}, \bibinfo{journal}{Proc. Natl. Acad. Sci. U.S.A.}
  \textbf{\bibinfo{volume}{106}}, \bibinfo{pages}{17640}
  (\bibinfo{year}{2009}).

\bibitem[{\citenamefont{Tse et~al.}(2007)\citenamefont{Tse, Yao, and
  Tanaka}}]{PhysRevLett.98.117004}
\bibinfo{author}{\bibfnamefont{J.~S.} \bibnamefont{Tse}},
  \bibinfo{author}{\bibfnamefont{Y.}~\bibnamefont{Yao}}, \bibnamefont{and}
  \bibinfo{author}{\bibfnamefont{K.}~\bibnamefont{Tanaka}},
  \bibinfo{journal}{Phys. Rev. Lett.} \textbf{\bibinfo{volume}{98}},
  \bibinfo{pages}{117004} (\bibinfo{year}{2007}).

\bibitem[{\citenamefont{Tang et~al.}(2009)\citenamefont{Tang, Sanville, and
  Henkelman}}]{JPhysCondensMatter.21.084204}
\bibinfo{author}{\bibfnamefont{W.}~\bibnamefont{Tang}},
  \bibinfo{author}{\bibfnamefont{E.}~\bibnamefont{Sanville}}, \bibnamefont{and}
  \bibinfo{author}{\bibfnamefont{G.}~\bibnamefont{Henkelman}},
  \bibinfo{journal}{J. Phys. Condens. Matter} \textbf{\bibinfo{volume}{21}},
  \bibinfo{pages}{084204} (\bibinfo{year}{2009}).

\bibitem[{\citenamefont{McMillan}(1968)}]{PhysRev.167.331}
\bibinfo{author}{\bibfnamefont{W.~L.} \bibnamefont{McMillan}},
  \bibinfo{journal}{Phys. Rev.} \textbf{\bibinfo{volume}{167}},
  \bibinfo{pages}{331} (\bibinfo{year}{1968}).

\bibitem[{\citenamefont{Chen et~al.}(1992)\citenamefont{Chen, Lewis, Su, Yu,
  and Cohen}}]{PhysRevB.46.5523}
\bibinfo{author}{\bibfnamefont{A.~L.} \bibnamefont{Chen}},
  \bibinfo{author}{\bibfnamefont{S.~P.} \bibnamefont{Lewis}},
  \bibinfo{author}{\bibfnamefont{Z.}~\bibnamefont{Su}},
  \bibinfo{author}{\bibfnamefont{P.~Y.} \bibnamefont{Yu}}, \bibnamefont{and}
  \bibinfo{author}{\bibfnamefont{M.~L.} \bibnamefont{Cohen}},
  \bibinfo{journal}{Phys. Rev. B} \textbf{\bibinfo{volume}{46}},
  \bibinfo{pages}{5523} (\bibinfo{year}{1992}).

\bibitem[{\citenamefont{{Berman} and {Brandt}}(1969)}]{JETPL.10.55B}
\bibinfo{author}{\bibfnamefont{I.~V.} \bibnamefont{{Berman}}} \bibnamefont{and}
  \bibinfo{author}{\bibfnamefont{N.~B.} \bibnamefont{{Brandt}}},
  \bibinfo{journal}{J. Exp. Theor. Phys. Lett.} \textbf{\bibinfo{volume}{10}},
  \bibinfo{pages}{55} (\bibinfo{year}{1969}).

\bibitem[{\citenamefont{Allen and Dynes}(1975)}]{PhysRevB.12.905}
\bibinfo{author}{\bibfnamefont{P.~B.} \bibnamefont{Allen}} \bibnamefont{and}
  \bibinfo{author}{\bibfnamefont{R.~C.} \bibnamefont{Dynes}},
  \bibinfo{journal}{Phys. Rev. B} \textbf{\bibinfo{volume}{12}},
  \bibinfo{pages}{905} (\bibinfo{year}{1975}).

\bibitem[{\citenamefont{Wang et~al.}(2015{\natexlab{a}})\citenamefont{Wang,
  Wang, Tse, Iitaka, and Ma}}]{ChemSci.6.522}
\bibinfo{author}{\bibfnamefont{Z.}~\bibnamefont{Wang}},
  \bibinfo{author}{\bibfnamefont{H.}~\bibnamefont{Wang}},
  \bibinfo{author}{\bibfnamefont{J.~S.} \bibnamefont{Tse}},
  \bibinfo{author}{\bibfnamefont{T.}~\bibnamefont{Iitaka}}, \bibnamefont{and}
  \bibinfo{author}{\bibfnamefont{Y.}~\bibnamefont{Ma}}, \bibinfo{journal}{Chem.
  Sci.} \textbf{\bibinfo{volume}{6}}, \bibinfo{pages}{522}
  (\bibinfo{year}{2015}{\natexlab{a}}).

\bibitem[{\citenamefont{Wang et~al.}(2015{\natexlab{b}})\citenamefont{Wang,
  Wang, Tse, Iitaka, and Ma}}]{PhysChemChemPhys.17.19379}
\bibinfo{author}{\bibfnamefont{Y.}~\bibnamefont{Wang}},
  \bibinfo{author}{\bibfnamefont{H.}~\bibnamefont{Wang}},
  \bibinfo{author}{\bibfnamefont{J.~S.} \bibnamefont{Tse}},
  \bibinfo{author}{\bibfnamefont{T.}~\bibnamefont{Iitaka}}, \bibnamefont{and}
  \bibinfo{author}{\bibfnamefont{Y.}~\bibnamefont{Ma}}, \bibinfo{journal}{Phys.
  Chem. Chem. Phys.} \textbf{\bibinfo{volume}{17}}, \bibinfo{pages}{19379}
  (\bibinfo{year}{2015}{\natexlab{b}}).

\bibitem[{\citenamefont{Slater}(1964)}]{JChemPhys.41.3199}
\bibinfo{author}{\bibfnamefont{J.~C.} \bibnamefont{Slater}},
  \bibinfo{journal}{J. Chem. Phys.} \textbf{\bibinfo{volume}{41}},
  \bibinfo{pages}{3199} (\bibinfo{year}{1964}).

\bibitem[{\citenamefont{Lide}(2004)}]{lide2004crc}
\bibinfo{author}{\bibfnamefont{D.~R.} \bibnamefont{Lide}},
  \emph{\bibinfo{title}{CRC handbook of chemistry and physics}}
  (\bibinfo{publisher}{CRC press}, \bibinfo{address}{Boca Raton},
  \bibinfo{year}{2004}).

\bibitem[{\citenamefont{Baettig and Zurek}(2011)}]{PhysRevLett.106.237002}
\bibinfo{author}{\bibfnamefont{P.}~\bibnamefont{Baettig}} \bibnamefont{and}
  \bibinfo{author}{\bibfnamefont{E.}~\bibnamefont{Zurek}},
  \bibinfo{journal}{Phys. Rev. Lett.} \textbf{\bibinfo{volume}{106}},
  \bibinfo{pages}{237002} (\bibinfo{year}{2011}).

\bibitem[{\citenamefont{Muramatsu et~al.}(2015)\citenamefont{Muramatsu, Wanene,
  Somayazulu, Vinitsky, Chandra, Strobel, Struzhkin, and
  Hemley}}]{JPhysChemC.119.18007}
\bibinfo{author}{\bibfnamefont{T.}~\bibnamefont{Muramatsu}},
  \bibinfo{author}{\bibfnamefont{W.~K.} \bibnamefont{Wanene}},
  \bibinfo{author}{\bibfnamefont{M.}~\bibnamefont{Somayazulu}},
  \bibinfo{author}{\bibfnamefont{E.}~\bibnamefont{Vinitsky}},
  \bibinfo{author}{\bibfnamefont{D.}~\bibnamefont{Chandra}},
  \bibinfo{author}{\bibfnamefont{T.~A.} \bibnamefont{Strobel}},
  \bibinfo{author}{\bibfnamefont{V.~V.} \bibnamefont{Struzhkin}},
  \bibnamefont{and} \bibinfo{author}{\bibfnamefont{R.~J.}
  \bibnamefont{Hemley}}, \bibinfo{journal}{J. Phys. Chem. C}
  \textbf{\bibinfo{volume}{119}}, \bibinfo{pages}{18007}
  (\bibinfo{year}{2015}).

\bibitem[{\citenamefont{Goncharenko et~al.}(2008)\citenamefont{Goncharenko,
  Eremets, Hanfland, Tse, Amboage, Yao, and Trojan}}]{PhysRevLett.100.045504}
\bibinfo{author}{\bibfnamefont{I.}~\bibnamefont{Goncharenko}},
  \bibinfo{author}{\bibfnamefont{M.~I.} \bibnamefont{Eremets}},
  \bibinfo{author}{\bibfnamefont{M.}~\bibnamefont{Hanfland}},
  \bibinfo{author}{\bibfnamefont{J.~S.} \bibnamefont{Tse}},
  \bibinfo{author}{\bibfnamefont{M.}~\bibnamefont{Amboage}},
  \bibinfo{author}{\bibfnamefont{Y.}~\bibnamefont{Yao}}, \bibnamefont{and}
  \bibinfo{author}{\bibfnamefont{I.~A.} \bibnamefont{Trojan}},
  \bibinfo{journal}{Phys. Rev. Lett.} \textbf{\bibinfo{volume}{100}},
  \bibinfo{pages}{045504} (\bibinfo{year}{2008}).

\bibitem[{\citenamefont{Eremets et~al.}(2003)\citenamefont{Eremets, Struzhkin,
  kwang Mao, and Hemley}}]{PhysicaBCondensedMatter.329.1312}
\bibinfo{author}{\bibfnamefont{M.~I.} \bibnamefont{Eremets}},
  \bibinfo{author}{\bibfnamefont{V.~V.} \bibnamefont{Struzhkin}},
  \bibinfo{author}{\bibfnamefont{H.}~\bibnamefont{kwang Mao}},
  \bibnamefont{and} \bibinfo{author}{\bibfnamefont{R.~J.}
  \bibnamefont{Hemley}}, \bibinfo{journal}{Physica B}
  \textbf{\bibinfo{volume}{329}}, \bibinfo{pages}{1312} (\bibinfo{year}{2003}).

\bibitem[{\citenamefont{Van~Eldik and Kl{\"a}rner}(2008)}]{van2008high}
\bibinfo{author}{\bibfnamefont{R.}~\bibnamefont{Van~Eldik}} \bibnamefont{and}
  \bibinfo{author}{\bibfnamefont{F.-G.} \bibnamefont{Kl{\"a}rner}},
  \emph{\bibinfo{title}{High pressure chemistry}} (\bibinfo{publisher}{John
  Wiley \& Sons}, \bibinfo{address}{Erlangen}, \bibinfo{year}{2008}).

\bibitem[{\citenamefont{Lee et~al.}(1992)\citenamefont{Lee, Vanderbilt,
  Laasonen, Car, and Parrinello}}]{PhysRevLett.69.462}
\bibinfo{author}{\bibfnamefont{C.}~\bibnamefont{Lee}},
  \bibinfo{author}{\bibfnamefont{D.}~\bibnamefont{Vanderbilt}},
  \bibinfo{author}{\bibfnamefont{K.}~\bibnamefont{Laasonen}},
  \bibinfo{author}{\bibfnamefont{R.}~\bibnamefont{Car}}, \bibnamefont{and}
  \bibinfo{author}{\bibfnamefont{M.}~\bibnamefont{Parrinello}},
  \bibinfo{journal}{Phys. Rev. Lett.} \textbf{\bibinfo{volume}{69}},
  \bibinfo{pages}{462} (\bibinfo{year}{1992}).

\bibitem[{\citenamefont{Aoki et~al.}(1996)\citenamefont{Aoki, Yamawaki,
  Sakashita, and Fujihisa}}]{PhysRevB.54.15673}
\bibinfo{author}{\bibfnamefont{K.}~\bibnamefont{Aoki}},
  \bibinfo{author}{\bibfnamefont{H.}~\bibnamefont{Yamawaki}},
  \bibinfo{author}{\bibfnamefont{M.}~\bibnamefont{Sakashita}},
  \bibnamefont{and} \bibinfo{author}{\bibfnamefont{H.}~\bibnamefont{Fujihisa}},
  \bibinfo{journal}{Phys. Rev. B} \textbf{\bibinfo{volume}{54}},
  \bibinfo{pages}{15673} (\bibinfo{year}{1996}).

\bibitem[{\citenamefont{Hirsch and Holzapfel}(1986)}]{JChemPhys.84.2771}
\bibinfo{author}{\bibfnamefont{K.~R.} \bibnamefont{Hirsch}} \bibnamefont{and}
  \bibinfo{author}{\bibfnamefont{W.~B.} \bibnamefont{Holzapfel}},
  \bibinfo{journal}{J. Chem. Phys.} \textbf{\bibinfo{volume}{84}},
  \bibinfo{pages}{2771} (\bibinfo{year}{1986}).

\bibitem[{\citenamefont{{Ge} et~al.}(2015)\citenamefont{{Ge}, {Zhang}, and
  {Yao}}}]{arXiv.1507.08525G}
\bibinfo{author}{\bibfnamefont{Y.}~\bibnamefont{{Ge}}},
  \bibinfo{author}{\bibfnamefont{F.}~\bibnamefont{{Zhang}}}, \bibnamefont{and}
  \bibinfo{author}{\bibfnamefont{Y.}~\bibnamefont{{Yao}}}, p.
  \bibinfo{pages}{arXiv:1507.08525} (\bibinfo{year}{2015}).

\bibitem[{\citenamefont{{Struzhkin} et~al.}(2014)\citenamefont{{Struzhkin},
  {Kim}, {Stavrou}, {Muramatsu}, {Mao}, {Pickard}, {Needs}, {Prakapenka}, and
  {Goncharov}}}]{arXiv.1412.1542S}
\bibinfo{author}{\bibfnamefont{V.~V.} \bibnamefont{{Struzhkin}}},
  \bibinfo{author}{\bibfnamefont{D.}~\bibnamefont{{Kim}}},
  \bibinfo{author}{\bibfnamefont{E.}~\bibnamefont{{Stavrou}}},
  \bibinfo{author}{\bibfnamefont{T.}~\bibnamefont{{Muramatsu}}},
  \bibinfo{author}{\bibfnamefont{H.}~\bibnamefont{{Mao}}},
  \bibinfo{author}{\bibfnamefont{C.~J.} \bibnamefont{{Pickard}}},
  \bibinfo{author}{\bibfnamefont{R.~J.} \bibnamefont{{Needs}}},
  \bibinfo{author}{\bibfnamefont{V.~B.} \bibnamefont{{Prakapenka}}},
  \bibnamefont{and} \bibinfo{author}{\bibfnamefont{A.~F.}
  \bibnamefont{{Goncharov}}}, p. \bibinfo{pages}{arXiv:1412.1542}
  (\bibinfo{year}{2014}).

\bibitem[{\citenamefont{{Drozdov} et~al.}(2015)\citenamefont{{Drozdov},
  {Eremets}, and {Troyan}}}]{arXiv.1508.06224D}
\bibinfo{author}{\bibfnamefont{A.~P.} \bibnamefont{{Drozdov}}},
  \bibinfo{author}{\bibfnamefont{M.~I.} \bibnamefont{{Eremets}}},
  \bibnamefont{and} \bibinfo{author}{\bibfnamefont{I.~A.}
  \bibnamefont{{Troyan}}}, p. \bibinfo{pages}{arXiv:1508.06224}
  (\bibinfo{year}{2015}).

\bibitem[{\citenamefont{{Shamp} et~al.}(2015)\citenamefont{{Shamp}, {Terpstra},
  {Bi}, {Falls}, {Avery}, and {Zurek}}}]{arXiv.1509.05455S}
\bibinfo{author}{\bibfnamefont{A.}~\bibnamefont{{Shamp}}},
  \bibinfo{author}{\bibfnamefont{T.}~\bibnamefont{{Terpstra}}},
  \bibinfo{author}{\bibfnamefont{T.}~\bibnamefont{{Bi}}},
  \bibinfo{author}{\bibfnamefont{Z.}~\bibnamefont{{Falls}}},
  \bibinfo{author}{\bibfnamefont{P.}~\bibnamefont{{Avery}}}, \bibnamefont{and}
  \bibinfo{author}{\bibfnamefont{E.}~\bibnamefont{{Zurek}}},
  \bibinfo{journal}{ArXiv e-prints} p. \bibinfo{pages}{arXiv:1509.05455}
  (\bibinfo{year}{2015}).

\bibitem[{\citenamefont{{Flores-Livas}
  et~al.}(2015{\natexlab{b}})\citenamefont{{Flores-Livas}, {Amsler}, {Heil},
  {Sanna}, {Boeri}, {Profeta}, {Wolverton}, {Goedecker}, and
  {Gross}}}]{arXiv.1512.02132F}
\bibinfo{author}{\bibfnamefont{J.~A.} \bibnamefont{{Flores-Livas}}},
  \bibinfo{author}{\bibfnamefont{M.}~\bibnamefont{{Amsler}}},
  \bibinfo{author}{\bibfnamefont{C.}~\bibnamefont{{Heil}}},
  \bibinfo{author}{\bibfnamefont{A.}~\bibnamefont{{Sanna}}},
  \bibinfo{author}{\bibfnamefont{L.}~\bibnamefont{{Boeri}}},
  \bibinfo{author}{\bibfnamefont{G.}~\bibnamefont{{Profeta}}},
  \bibinfo{author}{\bibfnamefont{C.}~\bibnamefont{{Wolverton}}},
  \bibinfo{author}{\bibfnamefont{S.}~\bibnamefont{{Goedecker}}},
  \bibnamefont{and} \bibinfo{author}{\bibfnamefont{E.~K.~U.}
  \bibnamefont{{Gross}}}, p. \bibinfo{pages}{arXiv:1512.02132}
  (\bibinfo{year}{2015}{\natexlab{b}}).

\bibitem[{\citenamefont{{Ma} et~al.}(2015)\citenamefont{{Ma}, {Duan}, {Li},
  {Liu}, {Tian}, {Huang}, {Zhao}, {Yu}, {Liu}, and {Cui}}}]{arXiv.1506.03889M}
\bibinfo{author}{\bibfnamefont{Y.}~\bibnamefont{{Ma}}},
  \bibinfo{author}{\bibfnamefont{D.}~\bibnamefont{{Duan}}},
  \bibinfo{author}{\bibfnamefont{D.}~\bibnamefont{{Li}}},
  \bibinfo{author}{\bibfnamefont{Y.}~\bibnamefont{{Liu}}},
  \bibinfo{author}{\bibfnamefont{F.}~\bibnamefont{{Tian}}},
  \bibinfo{author}{\bibfnamefont{X.}~\bibnamefont{{Huang}}},
  \bibinfo{author}{\bibfnamefont{Z.}~\bibnamefont{{Zhao}}},
  \bibinfo{author}{\bibfnamefont{H.}~\bibnamefont{{Yu}}},
  \bibinfo{author}{\bibfnamefont{B.}~\bibnamefont{{Liu}}}, \bibnamefont{and}
  \bibinfo{author}{\bibfnamefont{T.}~\bibnamefont{{Cui}}}, p.
  \bibinfo{pages}{arXiv:1506.03889} (\bibinfo{year}{2015}).

\bibitem[{\citenamefont{Oganov and Glass}(2006)}]{JChemPhys124.244704}
\bibinfo{author}{\bibfnamefont{A.~R.} \bibnamefont{Oganov}} \bibnamefont{and}
  \bibinfo{author}{\bibfnamefont{C.~W.} \bibnamefont{Glass}},
  \bibinfo{journal}{J. Chem. Phys.} \textbf{\bibinfo{volume}{124}},
  \bibinfo{pages}{244704} (\bibinfo{year}{2006}).

\end{thebibliography}

\end{document}


\title{\textbf{Supplemental Material} for ``High-pressure Phase Stability and Superconductivity of Pnictogen Hydrides and Chemical Trends for Compressed Hydrides''}

\author{Yuhao Fu}
\affiliation{College of Materials Science and Engineering and Key Laboratory of Automobile Materials of MOE, Jilin University, Changchun 130012, China}
\author{Xiangbo Du}
\affiliation{State Key Laboratory of Superhard Materials, Jilin University, Changchun 130012, China}
\author{Lijun Zhang}
\affiliation{College of Materials Science and Engineering and Key Laboratory of Automobile Materials of MOE, Jilin University, Changchun 130012, China}
\author{Feng Peng}
\affiliation{State Key Laboratory of Superhard Materials, Jilin University, Changchun 130012, China}
\author{Miao Zhang}
\affiliation{State Key Laboratory of Superhard Materials, Jilin University, Changchun 130012, China}
\author{Chris J. Pickard}
\affiliation{Department of Materials Science $\&$ Metallurgy, University of Cambridge, 27 Charles Babbage Road, Cambridge CB3 0FS, United Kingdom}
\author{Richard J. Needs}
\affiliation{Theory of Condensed Matter Group, Cavendish Laboratory, J J Thomson Avenue, Cambridge CB3 0HE, United Kingdom}
\author{David J. Singh}
\affiliation{College of Materials Science and Engineering and Key Laboratory of Automobile Materials of MOE, Jilin University, Changchun 130012, China}
\affiliation{Department of Physics and Astronomy, University of Missouri, Columbia, MO 65211-7010 USA}
\author{Weitao Zheng}
\affiliation{College of Materials Science and Engineering and Key Laboratory of Automobile Materials of MOE, Jilin University, Changchun 130012, China}
\author{Yanming Ma}
\affiliation{State Key Laboratory of Superhard Materials, Jilin University, Changchun 130012, China}


\date{\today}

\maketitle

\pagebreak
\clearpage
\vspace*{\fill}
\begin{figure}[h]
\includegraphics[width=6.5in]{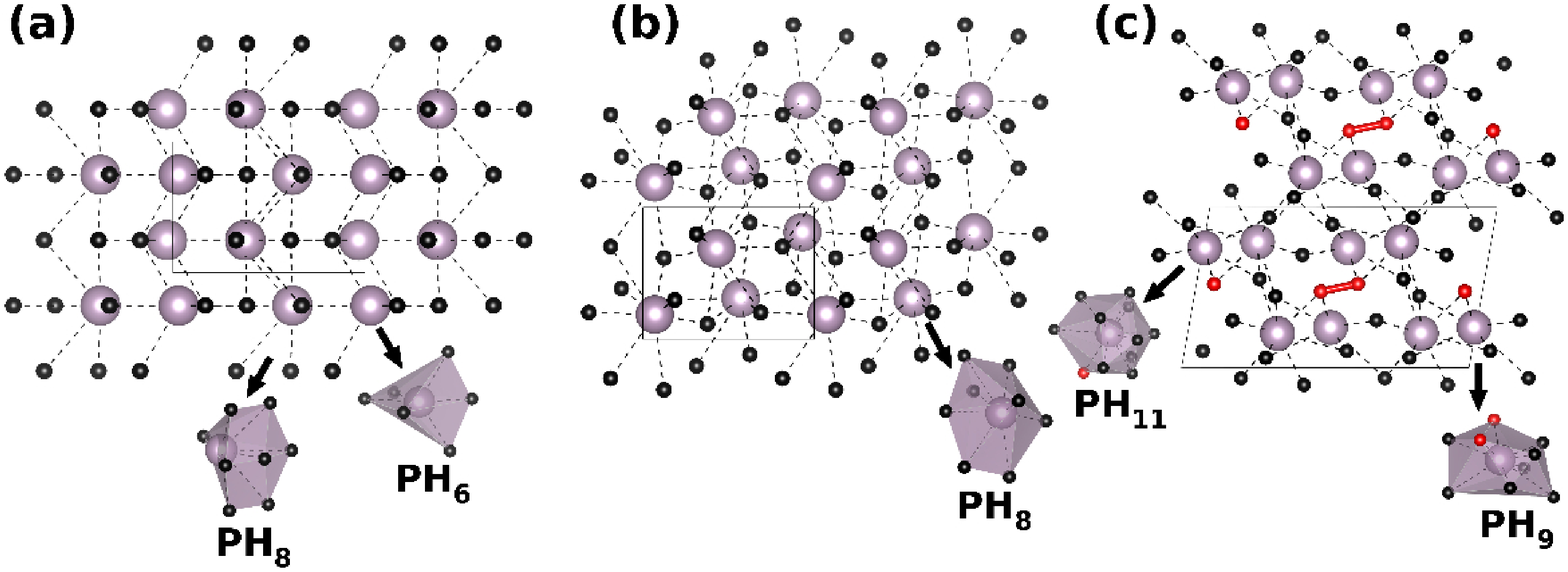}
\centering
\caption{(color online) Crystal structures of several predicted metastable P hydrides: (a) P$_2$H$_3$ (space group $P2_{1}/m$), (b) PH$_2$ space group $P2_{1}/c$), (c) PH$_3$ space group $C2/m$).}
\label{metaStructure}
\end{figure}
\vfill
\clearpage

\pagebreak
\clearpage
\vspace*{\fill}
\begin{figure}[h]
\begin{center}
\includegraphics[width=5in]{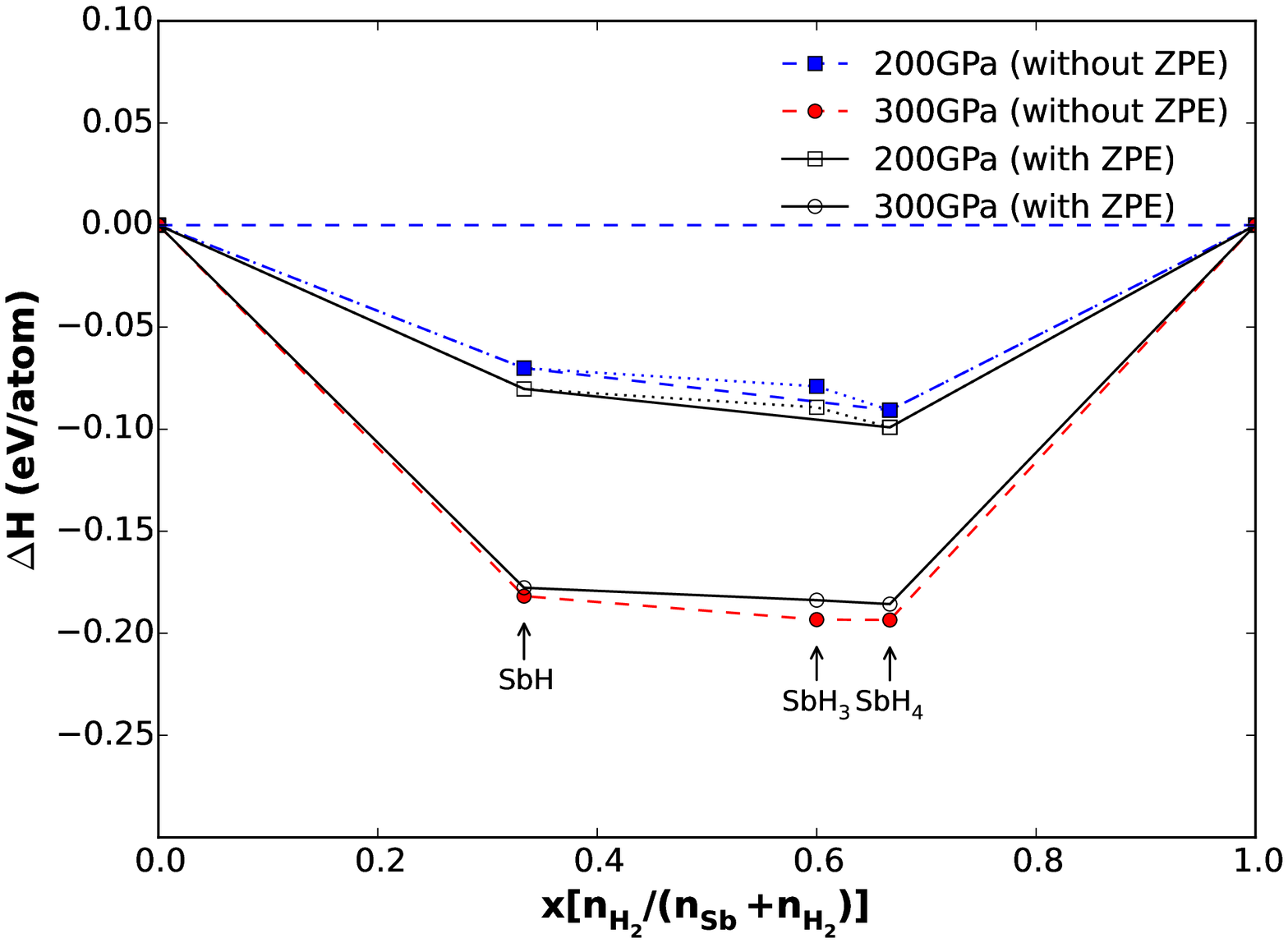}
\centering
\caption{(color online) Calculated formation enthalpy $\Delta$H (in
  meV/atom) of predicted stable Sb hydrides at 200 and 300 GPa with
  (open symbols) and without (filled symbols) inclusion of zero point energy (ZPE) effects, respectively.}
\label{convexhull-zpe}
\end{center}
\end{figure}
\vfill
\clearpage

\pagebreak
\clearpage
\vspace*{\fill}
\begin{figure}[h]
\begin{centering}
\includegraphics[width=5in]{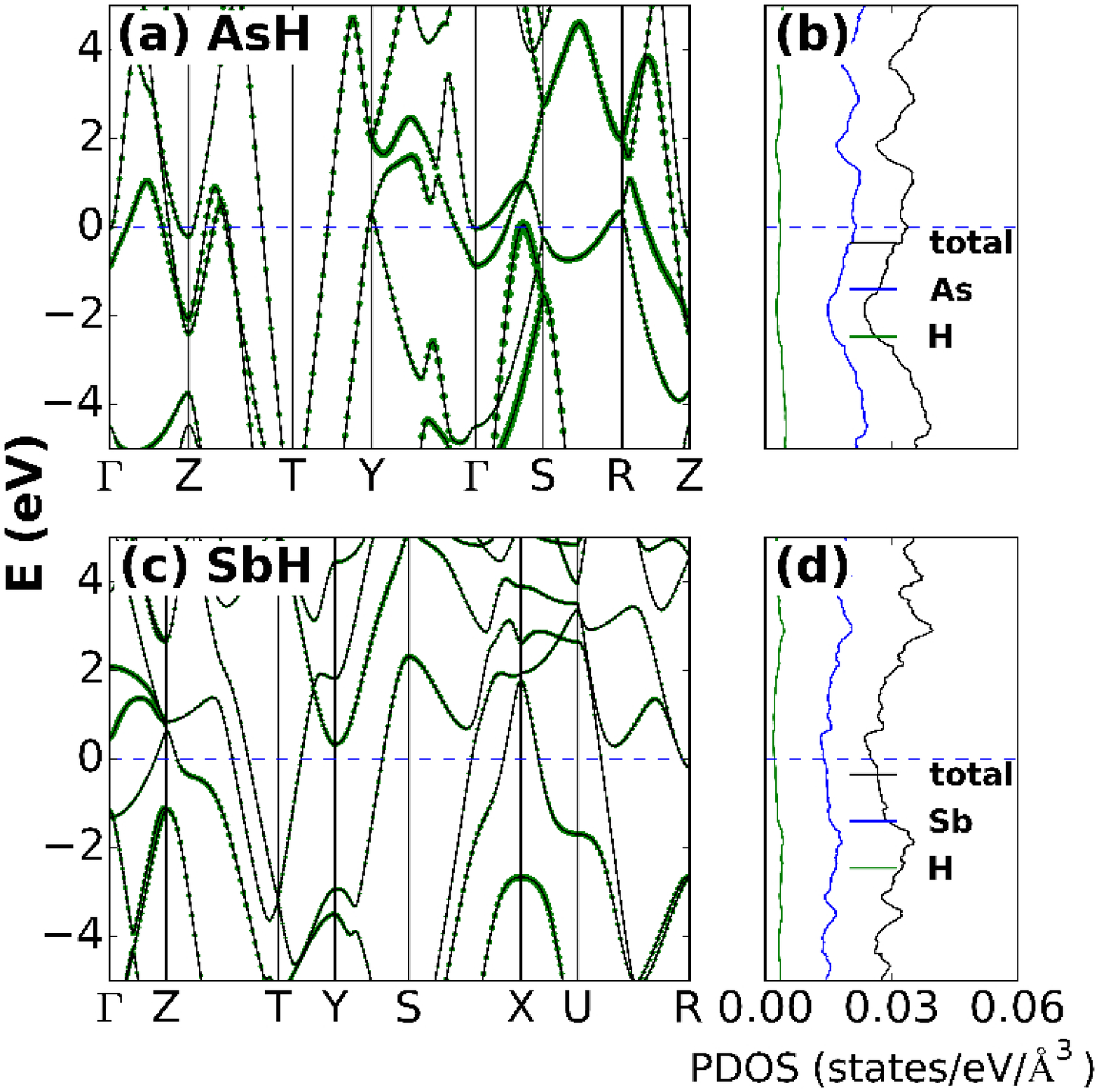}
\centering
\caption{(color online) Electronic band structure and projected
  density of states of (a,b) AsH at 350 GPa and (c,d) SbH at 300
  GPa. The green circles in the band structures represent the orbital
  projection of the electronic states onto H atoms.}
\label{stableBandStructure2}
\end{centering}
\end{figure}
\vfill
\clearpage

\pagebreak
\clearpage
\vspace*{\fill}
\begin{figure}[h]
\includegraphics[width=5in]{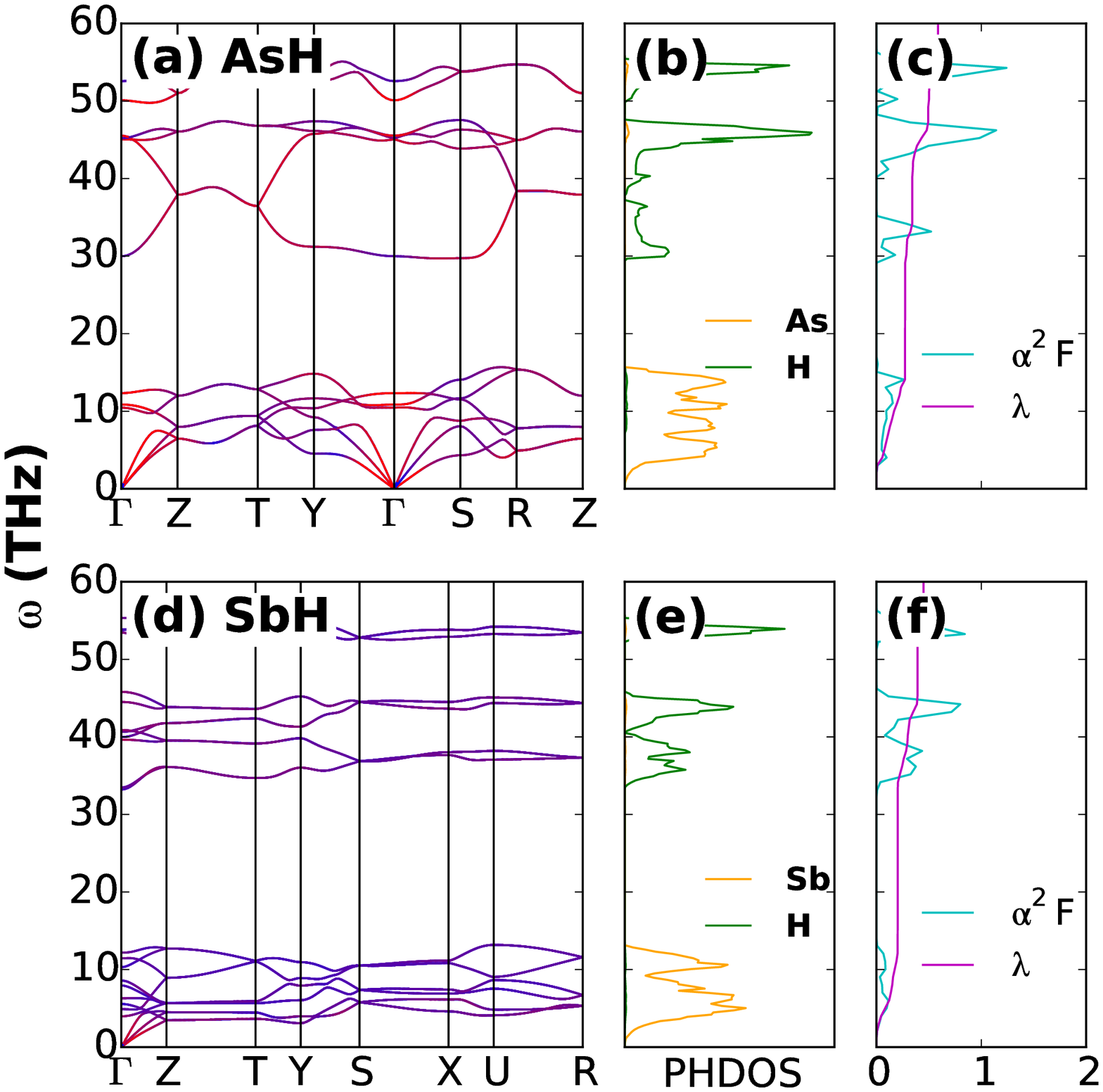}
\centering
\caption{(color online) Phonon dispersion curves (a,d), projected
  phonon densities of states (PHDOS) (b,e), Eliashberg EPC spectral
  functions $\alpha^{2}$F$(\omega)$ and its integral $\lambda(\omega)$
  (c,f) of AsH at 350 GPa and SbH at 300 GPa. The
  proportion of red color in the phonon dispersion curves represents
  the magnitude of the EPC parameter $\lambda_{n,q}(\omega)$ at each
  phonon mode ($n,q$).}
\label{stableDispersion2}
\end{figure}
\vfill
\clearpage

\pagebreak
\clearpage
\vspace*{\fill}
\begin{figure}[h]
\includegraphics[width=5in]{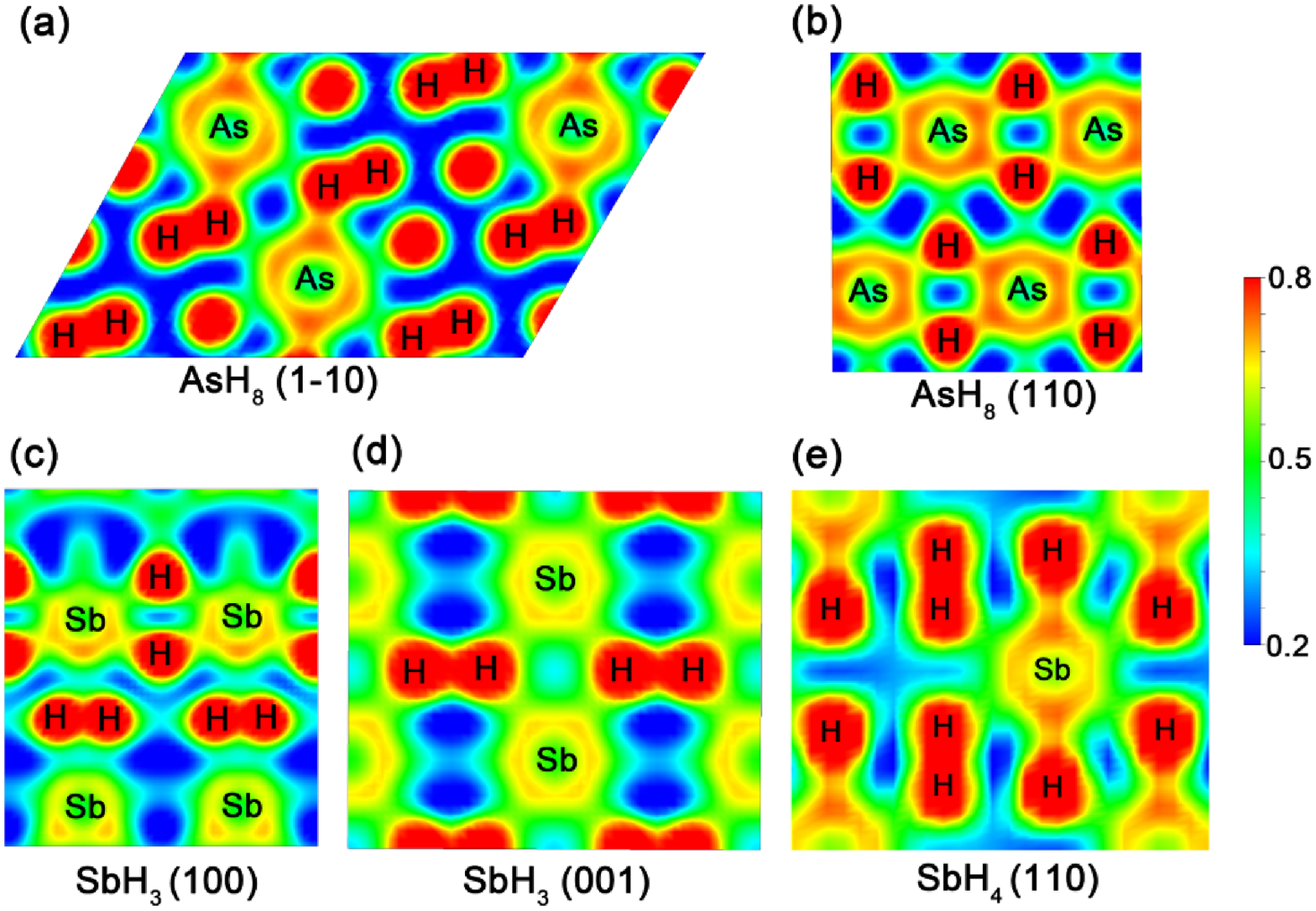}
\centering
\caption{(color online) Two-dimensional contour plots of the electron
  localization function for AsH$_{8}$ (a,b), SbH$_{3}$ (c,d) and
  SbH$_{4}$ (e), respectively. The planes containing As/Sb-H bonds are
  chosen to show the bonding character.}
\label{ELF}
\end{figure}
\vfill
\clearpage

\pagebreak
\clearpage
\vspace*{\fill}
\begin{table}[h]
\caption{Detailed structural information of several predicted metastable P hydrides.}
\centering
\resizebox{16cm}{!}{
\begin{tabular}{lclccl}
\hline\hline
\textbf{Phase (pressure)} & \textbf{Lattice parameters ($\AA$, $\circ$)} &  \multicolumn{4}{c}{\textbf{Atomic coordinates (fractional)}} \\
 \hline
P$_{2}$H$_{3}-P2_{1}/m$ (300GPa) & a=4.142 & \,\,\, P \, (2e) & \,\,\,   0.970 & \,\,\, 0.250 & \,\,\, 0.239 \\
                                    & b=2.825 & \,\,\, P \, (2e) & \,\,\, 0.377 & \,\,\, 0.250 & \,\,\, 0.657 \\
                                    & c=3.093 & \,\,\, H \, (2e) & \,\,\, 0.831 & \,\,\, 0.250 & \,\,\, 0.736 \\
                                    & $\alpha$=$\gamma$=90 & \,\,\, H \, (2e) & \,\,\, 0.386 & \,\,\, 0.750 & \,\,\, 0.002 \\
				    & $\beta$=89.838 & \,\,\, H \, (2e)  & \,\,\, 0.670 & \,\,\, 0.750 & 0.826 \\
\hline
PH$_{2}-P2_{1}/c$ (300GPa) & a=3.290 & \,\,\, P \, (4e) & \,\,\, 0.806 & \,\,\, 0.430 & \,\,\, 0.688 \\
                                              & b=3.920 & \,\,\, H \, (4e) & \,\,\, 0.253 & \,\,\, 0.187 & \,\,\, 0.297 \\
                                              & c=3.138 & \,\,\, H \, (4e) & \,\,\, 0.477 & \,\,\, 0.876 & \,\,\, 0.384 \\
                                              & $\alpha$=$\gamma$=90 &  &  &  &  \\
                                              & $\beta$=106.456 &  &  &  &  \\
\hline
PH$_{3}-C2/m$ (300GPa) & a=7.206 & \,\,\, P \, (4i) & \,\,\, 0.993 & \,\,\, 0.000 & \,\,\, 0.748 \\
                                        & b=3.093 & \,\,\, P \, (4i) & \,\,\, 0.813 & \,\,\, 0.500 & \,\,\, 0.785 \\
                                        & c=4.088 & \,\,\, H \, (8j) & \,\,\, 0.852 & \,\,\, 0.296 & \,\,\, 0.296 \\
                                        & $\alpha$=$\gamma$=90 & \,\,\, H \, (4i) & \,\,\, 0.213 & \,\,\, 0.000 & \,\,\, 0.064 \\
					& $\beta$=100.029 & \,\,\, H \, (4i) & \,\,\, 0.783 & \,\,\, 0.000 & \,\,\, 0.554 \\
					&  & \,\,\, H \, (4i) & \,\,\, 0.936 & \,\,\, 0.500 & \,\,\, 0.106 \\
					&  & \,\,\, H \, (4i) & \,\,\, 0.062 & \,\,\, 0.500 & \,\,\, 0.477 \\
\hline
\end{tabular}
  }
\label{structure3}
\end{table}
\vfill
\clearpage

\pagebreak
\clearpage
\vspace*{\fill}
\begin{table}[h]
\caption{Detailed structural information on the predicted stable pnictogen hydrides.}
\centering
\resizebox{16cm}{!}{
\begin{tabular}{lclccl}
\hline\hline
\textbf{Phase (pressure)} & \textbf{Lattice parameters ($\AA$, $\circ$)} &  \multicolumn{4}{c}{\textbf{Atomic coordinates (fractional)}} \\
 \hline
AsH-$Cmcm$ (350GPa) & a=3.241 & \,\,\, As (4c) & \,\,\,   0.500 & \,\,\, 0.695 & \,\,\, 0.250 \\
                                    & b=4.156 & \,\,\, H \, (4c) & \,\,\, 0.500 & \,\,\, 0.932 & \,\,\, 0.750 \\
                                    & c=2.998 &        &            &             & \\
                                    & $\alpha$=$\beta$=$\gamma$=90 & & & & \\
\hline
AsH$_{8}-C2/c$ (350GPa) & a=5.604 & \,\,\, As (4e) & \,\,\, 0.000 & \,\,\, 0.246 & \,\,\, 0.750 \\
                                              & b=2.813 & \,\,\, H \, (8f) & \,\,\, 0.263 & \,\,\, 0.397 & \,\,\, 0.394 \\
                                              & c=5.685 & \,\,\, H \, (8f) & \,\,\, 0.771 & \,\,\, 0.387 & \,\,\, 0.871 \\
                                              & $\alpha$=$\gamma$=90 & \,\,\, H \, (8f) & \,\,\, 0.131 & \,\,\, 0.253 & \,\,\, 0.072 \\
                                              & $\beta$=120.033 & \,\,\, H \, (8f) & \,\,\, 0.993 & \,\,\, 0.753 & \,\,\, 0.603 \\
\hline
SbH$-Pnma$ (300GPa) & a=2.771 & \,\,\, Sb (4c) & \,\,\, 0.356 & \,\,\, 0.750 & \,\,\, 0.212 \\
                                        & b=3.197 & \,\,\, H \, (4c) & \,\,\, 0.925 & \,\,\, 0.750 & \,\,\, 0.855 \\
                                        & c=3.9852 & & & & \\
                                        & $\alpha$=$\beta$=$\gamma$=90 & & & & \\
\hline
SbH$_{3}-Pmmn$ (300GPa) & a=2.771 & \,\,\, Sb (2b) & \,\,\, 0.000 & \,\,\, 0.500 & \,\,\, 0.101 \\
                                                & b=3.197 & \,\,\, Sb (2b) & \,\,\, 0.000 & \,\,\, 0.500 & \,\,\, 0.631 \\
                                                & c=7.284 & \,\,\, H \, (4f) & \,\,\, 0.738 & \,\,\, 0.000 & \,\,\, 0.150 \\
                                                & $\alpha$=$\beta$=$\gamma$=90 & \,\,\, H \, (4e) & \,\,\, 0.500 & \,\,\, 0.143 & \,\,\, 0.642 \\
                                                & & \,\,\, H \, (2a) & \,\,\, 0.000 & \,\,\, 0.000 & \,\,\, 0.535 \\
                                                & & \,\,\, H \, (2a) & \,\,\, 0.000 & \,\,\, 0.000 & \,\,\, 0.748 \\
\hline
SbH$_{4}-P6_{3}/mmc$ (300GPa) & a=2.783 & \,\,\, Sb (2c) & \,\,\, 0.667 & \,\,\, 0.333 & \,\,\, 0.750 \\
                                                           & b=2.783 & \,\,\, H \, (4e) & \,\,\, 0.000 & \,\,\, 0.000 & \,\,\, 0.580 \\
                                                           & c=5.217 & \,\,\, H \, (4f) & \,\,\, 0.333 & \,\,\, 0.667 & \,\,\, 0.581 \\
                                                           & $\alpha$=$\beta$=90 & & & & \\
                                                           & $\gamma$=120 & & & & \\
\hline
\end{tabular}
  }
\label{structure2}
\end{table}
\vfill
\clearpage

\pagebreak
\clearpage
\vspace*{\fill}
\begin{table}[h]
  \caption{Bader charge analysis for the predicted stable pnictogen hydrides. 
    Note that H represents monoatomic H atoms bonding with As/Sb, and H1, H2, H3 and H4
    are the H atoms forming quasi-molecular H$_{2}$-units.}
\centering
\begin{tabular}{llcc}
\hline\hline
 &  &  \textbf{population} & \textbf{charge} \\
 \hline
AsH$-Cmcm$ & As & 4.6885 & 0.3115 \\
                        & H  & 1.3115 & -0.3115 \\
\hline
AsH$_{8}$ & As & 4.1217 & 0.8783 \\
                   & H1 & 1.1524 & -0.1524 \\
                   & H2 & 1.1248 & -0.1248 \\
                   & H3 & 1.0866 & -0.0866 \\
                   & H4 & 1.0753 & -0.0753 \\
\hline
SbH$-Pnma$ & Sb & 4.4666 & 0.5334 \\
                       & H  & 1.5334 & -0.5334 \\
\hline
SbH$_{3}-Pmmn$ & Sb1 & 3.9782 & 1.0218 \\
                               & Sb2 & 3.5377 & 1.4623 \\
                               & H1  & 1.3310 & -0.3310 \\
                               & H    & 1.4354 & -0.4354 \\
                               & H    & 1.4729 & -0.4749 \\

\hline
SbH$_{4}-P6_{3}/mmc$ & Sb & 3.5117 & 1.4883 \\
                                         & H1 & 1.2713 & -0.2713 \\
                                         & H   &  0.1430 & -0.4730 \\
\hline
\end{tabular}
\label{bader}
\end{table}
\vfill
\clearpage

\pagebreak
\clearpage
\vspace*{\fill}
\begin{table}[h]
  \caption{Calculated total EPC parameter $\lambda$, logarithmic average frequency 
    $\omega_{log}$, electronic DOS at the Fermi level $N(E_{f})$ and superconducting 
    critical temperature $T_{c}$ for the predicted stable pnictogen hydrides. The screened 
    Coulomb repulsion parameter $\mu^{*}=0.1$ is used for the $T_{c}$ calculations.}
\centering
\begin{tabular}{lccccc}
\hline\hline
 \textbf{Phase} & \textbf{Pressure\ (GPa)} &  \textbf{$\lambda$} & \textbf{$\omega_{log}\ (K)$}  & \textbf{$N(E_{f})$} & \textbf{$T_{c}\ (K)$}\\
 \hline
AsH$-Cmcm$ & 300 & 0.59 & 995 & 4.43 & 21.2 \\
                        & 400 & 0.55 & 1166 & 4.24 & 20.2 \\
\hline
AsH$_{8}-C2/c$ & 350 & 1.50 & 1240 & 7.29 & 141.0 \\
                            & 400 & 1.63 & 1169 & 7.44 & 143.9 \\
                            & 450 & 1.77 & 1153 & 7.56 & 151.4 \\
\hline
SbH$-Pnma$ & 175 & 0.57 & 736 & 10.40 & 14.6 \\
                       & 215 & 0.51 & 820 & 9.89 & 10.5 \\
                       & 255 & 0.47 & 877 & 9.59 & 8.5 \\
                       & 295 & 0.45 & 890 & 9.30 & 6.8 \\
\hline
SbH$_{3}-Pmmn$ & 300 & 0.61 & 1100 & 9.25 & 25.9 \\
                                & 400 & 0.56 & 1112 & 8.89 & 19.8 \\
\hline
SbH$_{4}-P6_{3}/mmc$ & 150 & 1.28 & 1054 & 8.07 & 102.2 \\
                                         & 200 & 1.21 & 1133 & 7.68 & 102.3 \\
                                         & 250 & 1.18 & 1145 & 7.38 & 99.9 \\
                                         & 300 & 1.16 & 1090 & 7.12 & 93.9 \\
\hline
\end{tabular}
\label{epc2}
\end{table}
\vfill
\clearpage